\documentclass[a4paper]{article}

\usepackage{a4wide}
\usepackage[UKenglish]{babel}
\usepackage{graphicx,subfigure}
    \graphicspath{{./}{figure/}{figure/gambling/}{figure/taxation/}}
\usepackage[colorlinks=true,linkcolor=blue,citecolor=red]{hyperref}
\usepackage{amsthm,mathtools,bbold}
\usepackage[inline]{enumitem}

\DeclarePairedDelimiter\ave{\langle}{\rangle}
\newcommand{\C}{\mathbb{C}}
\newcommand{\ff}{\hat{f}}

\renewcommand{\P}{\mathbb{P}}
\newcommand{\pr}[1]{{}^\ast\!#1}
\newcommand{\R}{\mathbb{R}}
\newcommand{\Var}{\operatorname{Var}}
\newcommand{\Z}{\mathbb{Z}}

\theoremstyle{remark}\newtheorem{remark}{Remark}

\begin{document}
\title{Kinetic modelling of multiple interactions \\ in socio-economic systems}

\author{Giuseppe Toscani\thanks{Department of Mathematics ``F. Casorati'', University of Pavia, Via Ferrata 1, 27100 Pavia, Italy} \and
        Andrea Tosin\thanks{Department of Mathematical Sciences ``G. L. Lagrange'', Politecnico di Torino, Corso Duca degli Abruzzi 24, 10129 Torino, Italy} \and
        Mattia Zanella\thanks{Department of Mathematical Sciences ``G. L. Lagrange'', Politecnico di Torino, Corso Duca degli Abruzzi 24, 10129 Torino, Italy}
       }
\date{}

\maketitle

\begin{abstract}
Unlike the classical kinetic theory of rarefied gases, where microscopic interactions among gas molecules are described as binary collisions, the modelling of socio-economic phenomena in a multi-agent system naturally requires to consider, in various situations, multiple interactions among the individuals. In this paper, we collect and discuss some examples related to economic and gambling activities. In particular, we focus on a linearisation strategy of the multiple interactions, which greatly simplifies the kinetic description of such systems while maintaining all their essential aggregate features, including the equilibrium distributions.

\medskip


\end{abstract}

\section{Introduction}
Unlike the classical kinetic theory of rarefied gases, where binary collisions are dominant, socio-economic phenomena are often characterised by simultaneous interactions among a large number $N\gg 1$ of individuals, which lead to highly non-linear Boltzmann-type equations for the evolution of the distribution function of the agents. The study of Boltzmann-type equations in presence of multiple interactions is quite recent. The theoretical approach to generalised Maxwell models with multiple interactions has been studied in~\cite{bobylev2009CMP} by resorting to the Fourier-transformed version of the kinetic equation, like in the case of the classical binary Maxwell-type interactions. In that paper, existence and uniqueness of solutions to the initial value problem were studied together with the large time behaviour, both in the case of convergence to a stationary state or of convergence to a self-similar solution.

The prototype of the kinetic models considered in~\cite{bobylev2009CMP} may be briefly described as follows. Given a certain number $N\gg 1$ of identical particles with pre-collisional state $V=(v_1,\,v_2,\,\dots,\,v_N)$, the $N$-particle interaction is understood as a random linear transformation of $V$ into the post-collisional state $V^\ast=(v_1^\ast,\,v_2^\ast,\,\dots,\,v_N^\ast)$, where
\begin{equation}
    v_i^\ast=av_i+b\sum_{j=1}^{N}v_j \qquad i=1,\,\dots,\,N,
    \label{eq:pos}
\end{equation}
and the parameters $a,\,b\in\R$ may be fixed or randomly distributed with a certain number of moments bounded. The transformation~\eqref{eq:pos} is nothing but a generalisation of a binary collision, which is recovered in the particular case $N=2$. Denoting by $f=f(v,\,t)$ the density of particles with state $v\in\R$ at time $t\geq 0$, and postulating the validity of molecular chaos, the evolution of any \textit{observable quantity} $\varphi$, i.e. any quantity which may be expressed as a function of $v$, is given by the Boltzmann-type equation
\begin{equation}
    \frac{d}{dt}\int_\R\varphi(v)f(v,\,t)\,dv=
        \frac{1}{\tau N}\int_{\R^N}\sum_{i=1}^{N}\ave*{\varphi(v_i^\ast)-\varphi(v_i)}\prod_{i=1}^{N}f(v_i,\,t)\,dv_1\,\dots\,dv_N,
    \label{eq:kine-Gamba}
\end{equation}
where $\tau$ denotes a relaxation time and $\ave{\cdot}$ is the average with respect to the distributions of the random parameters $a,\,b$ contained in~\eqref{eq:pos}. The collision integral on the right-hand side of~\eqref{eq:kine-Gamba} takes into account the whole set of microscopic states, consequently it depends on the $N$-product of the density functions $f(v_1,\,t)\cdot\ldots\cdot f(v_N,\,t)$. Thus, the evolution of $f$ obeys a highly non-linear Boltzmann-type equation. On the other hand, the collision integral features a constant \textit{collision kernel}, chosen equal to $1$ without loss of generality. This corresponds, in the jargon of the classical kinetic theory, to consider \textit{Maxwellian interactions}.

Remarkably, in~\cite{bobylev2009CMP} the main suggested application of~\eqref{eq:kine-Gamba} is in an economic context. Specifically, the interacting particles are considered as a community of agents participating in various economical trades and $V$ is the vector of their non-negative wealth. The post-collisional state $V^\ast$ gives the new wealth of the community of agents after a single economic trade. Unlike the classical interpretation of $V$ as a vector of velocities, which are defined on the whole space $\R$, in economic applications each $v_i$ takes values in $\R_+$, thus the domain of integration in~\eqref{eq:kine-Gamba} has to be changed accordingly.

A further important fact noticed in~\cite{bobylev2009CMP} is that, under the additional assumption of a very large number of interacting agents ($N\to\infty$), the economic game described by~\eqref{eq:pos} may be suitably linearised and the Boltzmann-type equation~\eqref{eq:kine-Gamba} becomes linear as well. Taking advantage of such a linearisation strategy, a particular application to economy of the multiple-interaction kinetic setting has then been studied in detail in~\cite{bobylev2011KRM}. There, a linearised model of a multiple-trading activity is derived in the limit of a large number of traders, whose behaviour reproduces, among other properties, the formation of power-like tails in the wealth distribution for large times.

More recently, the Maxwellian description introduced in~\cite{bobylev2009CMP} has been applied to study the jackpot, an online lottery-type game which occupies a big portion of the gambling market on the web~\cite{toscani2019PRE}. The large number of gamblers taking part in each round of the game allows one to treat them as a particular multi-agent economic system reminiscent of the kinetic description given in~\cite{bobylev2009CMP}. In this case, the $N$-particle interaction consists in a generalisation of~\eqref{eq:pos}:
\begin{equation}
    v_i^\ast= av_i+b_i\sum_{j=1}^{N}v_j+cY_i, \qquad i=1,\,\dots,\,N.
    \label{eq:jack}
\end{equation}
Here, $a=1-\epsilon>0$ and $c=\epsilon\beta>0$ are fixed constants depending on a small parameter $0<\epsilon\ll 1$. Conversely, the $B_i$'s are random parameters of the form $b_i=\epsilon(1-\delta)I(A-i)$ with $0<\delta <1$ and $I$ denoting a function which takes the value $1$ if its argument vanishes and the value $0$ otherwise. Finally, $A\in\{1,\,\dots,\,N\}$ is a discrete random variable giving the index of the gambler who wins in a particular round. The quantity $v_i\geq 0$ in~\eqref{eq:jack} represents the number of tickets, hence the amount of money, put into the game by the $i$th gambler while $v_i^\ast$ is the new number of tickets, i.e. the new amount of money, owned by the $i$th gambler after the draw of the winning ticket. Two main differences with respect to~\eqref{eq:pos} need to be stressed. First, the post-collisional state $V^\ast$ does not only depend on the pre-collisional state $V$ but also on $Y=(Y_1,\,Y_2,\,\dots,\,Y_N)$, where each $Y_i\geq 0$ is a random variable describing the refilling of tickets by gamblers to compensate for possible losses. Second, the random parameters $b_i$ depend on the gambler index $i$ through the pre-collisional state $V$. Owing to the definition of the random variable $I(A-i)$, it is easy to realise that, in each round, only one gambler will end up with the whole amount of money put into the game in that round.

A third significant example of economic application of the multiple-interaction setting comes from taxation and redistribution. In this case, the Boltzmann-type equation~\eqref{eq:kine-Gamba} describes the evolution of the wealth density produced by trades in which a percent amount of the invested wealth is taxed and simultaneously redistributed among all the traders. Here, the multiple interaction takes the form
$$ v_i^\ast=a_iv_i+\sum_{j=1}^{N}b_jv_j+cw_i, \qquad i=1,\,\dots,\,N, $$
with
$$ a_i=1-(1+\alpha)\lambda-(1-\alpha)\eta_i, \qquad b_i=\frac\alpha{N}(\lambda+\eta_i). $$
Specifically, $\lambda\in (0,\,1]$ is the fixed percent amount of wealth that the $i$th agent invests in the transaction, $\eta_i\in\R$ is a random variable modelling the intrinsic risk of the transaction and $\alpha\in [0,\,1]$ is the rate of taxation. Last, $c=\lambda$ and $w_i\geq 0$ is a random variable which quantifies the gain returned by the trade to the $i$th trader. The main difference with respect to~\eqref{eq:pos} is the presence of $N$ different random variables $b_i$. 

The previous examples share a common modelling feature: every interaction involves a certain fixed number $N$ of individuals, typically such that $N>2$. Then the nature of the interaction, including in particular the number of the involved individuals, determines the properties of the large time equilibrium or self-similar density, i.e. the analogous of the Maxwellian distribution in the classical kinetic theory, which depicts the emerging aggregate characteristics of the system. In general, it is difficult to extract precise information on $f$ from the highly non-linear kinetic equation~\eqref{eq:kine-Gamba}. Nevertheless, in all the considered cases if the number of interacting individuals is large enough then, in the limit $N\to\infty$, a linearisation of both the multiple interaction rule and the Boltzmann-type equation is possible, which makes the problem more amenable to analytical investigation. 

Remarkably, such an asymptotic regime, whose mathematical derivation is actually only formal, turns out to capture a quite detailed representation of the large time trend of the solution to~\eqref{eq:kine-Gamba}. In particular, numerical simulations of the ``real'' $N$-particle interactions clearly indicate that the solution of the linearised kinetic model is in very good agreement with that of~\eqref{eq:kine-Gamba} even for a moderate number of interacting agents, say $N=O(10^2)$, cf.~\cite{toscani2019PRE}.

In the rest of the paper we will present the main results about the three aforementioned examples of economic games and their asymptotic linearisation. While the first two examples, here summarised respectively in Sections~\ref{sect:Bobylev},~\ref{sect:gambling}, may be found in full detail in the original papers~\cite{bobylev2011KRM,toscani2019PRE}, the third example is new and is exhaustively discussed in Section~\ref{sect:taxation}. Numerical tests are finally presented in Section~\ref{sect:numerics}. 

For the sake of completeness, we remark that related examples of linearisation of economical interactions have been proposed in~\cite{toscani2013JSP} in connection with the problem of price formation in a multi-agent society in which agents interact by exchanging two types of goods. The results in~\cite{toscani2013JSP} have been subsequently generalised in~\cite{brugna2018PHYSA} to a society in which one has the simultaneous presence of two classes of agents: dealers and speculators. Both classes adopt the same strategy, but speculators play on quantities of goods to be exchanged to have a better return. However, in these examples the density function depends on two variables, and it does not fall in the description of~\cite{bobylev2009CMP}.

\section{Maxwell-type models in socio-economic problems}
We start by summarising the hallmarks of the application of the collisional kinetic theory to socio-economic problems.

Let us consider a large community of agents, who aim to improve their wealth condition by interacting with each other. As usual in the kinetic description, we assume that the agents are indistinguishable~\cite{pareschi2013BOOK}. This means that at any time $t\geq 0$ the state of a generic representative agent is completely characterised by their wealth, which is expressed by the amount $v\in\R_+$ of owned money. Consequently, the state of the community of agents can be described by a distribution function $f=f(v,\,t)$. The precise meaning of $f$ is the following: given a sub-domain $D\subseteq\R_+$, the quantity
$$ \int_Df(v,\,t)\,dv $$
is proportional to the number of individuals possessing a wealth $v\in D$ at time $t\geq 0$. If one assumes the following normalisation condition:
$$ \int_{\R_+}f(v,\,t)\,dv=1, $$
then $f$ may be conveniently understood as a probability density, hence $\int_Df(v,\,t)\,dv$ becomes the probability that at time $t\geq 0$ the wealth of a representative agent of the system belongs to $D$.

The time evolution of the density $f$ is due to the fact that the community of agents performs economical trades, which we also call \textit{games} to stress the difference with the classical \textit{collisions} of gas particles. The games happen at regular or random time intervals and involve, in general, a certain random number of agents, that continuously upgrade their wealth at each new trade. In analogy with the classical kinetic theory of rarefied gases, we refer to a single upgrade of the quantity $v$ as an \textit{interaction}. In order to avoid inessential difficulties, we may assume that the number of agents involved in an interaction is actually fixed, say $N\gg 1$. Then, once the details of the microscopic interaction have been prescribed, the evolution of the density $f$ obeys the Boltzmann-type equation~\eqref{eq:kine-Gamba}, which is a generalisation of the bilinear Boltzmann equation for Maxwell pseudo-molecules. As already mentioned in the Introduction, Maxwell-type agents are such that their interaction kernel, which in essence models the frequency of the interactions, is constant. In particular, it does not depend on the pre-interaction states of the agents.

An interesting motivation for the choice of a constant interaction kernel in socio-economic applications is provided in~\cite{bobylev2009CMP}. Maxwell models satisfy the very strong condition of scaling invariance, which implies that the time dynamics predicted by the model are invariant under the transformation $v\to\nu v$, where $\nu$ is a constant. In the economic context, this property means that the agents do not change their way of trading if they move to a new currency. Moreover, a constant interaction kernel also translates the realistic idea that the various amounts of money put into the game by different agents are uncorrelated. As we will see, the forthcoming examples fit perfectly into this assumption.

From the mathematical point of view, the choice of a constant interaction kernel has the further property that, by choosing $\varphi(v)=e^{-i\xi v}$ in~\eqref{eq:kine-Gamba}, where $i$ is the imaginary unit and $\xi\in\R$, and $v_i^\ast$ like in~\eqref{eq:pos}, one obtains a closed form for the Fourier-transformed Boltzmann-type equation, which reads:
$$ \partial_t\ff(\xi,\,t)=\frac{1}{N\tau}\ave*{\ff((A+B)\xi,\,t)\ff(B\xi,\,t)^{N-1}}-\frac{1}{\tau}\ff(\xi,\,t), $$
where, as usual, $\ff$ denotes the Fourier transform of the distribution function $f$:
$$ \ff(\xi,\,t):=\int_{\R_+}f(v,\,t)e^{-i\xi v}\,dv. $$ 
Note that for socio-economic problems, in which the microscopic state $v$ takes only non-negative values, one may also resort to the Laplace-transformed Boltzmann-type equation by choosing $\varphi(v)=e^{-sv}$ in~\eqref{eq:kine-Gamba} with either $s\in\R$ or $s\in\C$.

\section{Economic trades with a large number of traders}
\label{sect:Bobylev}
In~\cite{bobylev2011KRM}, the authors study an interesting economic application of the general theory of Maxwell models with multiple collisions discussed in~\cite{bobylev2009CMP}. They consider a community of individuals participating in a collective trade subject to a certain number of clearly identified conditions. In the original formulation, each trade involves a random number $N\geq 2$ of randomly chosen traders. Here, for the sake of simplicity, we consider $N\gg 1$ fixed and we denote by $v_i\geq 0$, $i=1,\,\dots,\,N$, the wealth of the $i$th trader before the trade. The aforementioned conditions of the trade are the following:
\begin{enumerate}[label=(\arabic*)]
\item the traders form a sum (total capital)
$$ S:=\sum_{i=1}^Nv_i, $$
and proceed to trade it;
\item the new sum $S^\ast$ resulting from the trade is obtained by multiplying $S$ by a random number $\theta\geq 0$ distributed according to a given probability density $g:\R_+\to\R_+$:
$$ S^\ast=\theta S=\sum_{i=1}^{N}v_i^\ast, $$
where $v_i^\ast:=\theta v_i$ for each $i$;
\item the sum $S^\ast$ is returned to the traders in accordance to the following rule: given a fixed or random number $\gamma\in (0,\,1)$, the part $S^\ast_1:=(1-\gamma)S^\ast$ is split among the traders proportionally to their initial contributions, whereas the remaining part $S^\ast_2:=\gamma S^\ast$ is distributed among all the traders equally. Hence, the trade is completely defined in terms of the probability density $g$ and the parameter $\gamma$.
\end{enumerate}

In~\cite{bobylev2011KRM}, the qualitative analysis of the kinetic model resulting from these microscopic rules is confined to the simple case in which the random number $\theta$ assumes only two fixed values, in particular $\theta=0,\,s>0$ with $\P(\theta=0)=q\in [0,\,1]$ and $\P(\theta=s)=1-q$. Therefore, the distribution $g$ takes the form
\begin{equation}
    g(\theta)=q\delta(\theta)+(1-q)\delta(\theta-s),
    \label{eq:gt}
\end{equation}
where $\delta(\theta)$ denotes the Dirac delta centred at $\theta=0$. Consequently, the outcome of the game depends on the three positive parameters $\gamma$, $s$, $q$.

The trade just described is an example of the multiple interaction setting considered in~\cite{bobylev2009CMP}. Indeed, it can be equivalently reformulated as a linear transformation of the pre-interaction wealth vector $(v_1,\,v_2,\,\dots,\,v_N)$ into the post-interaction wealth vector $(v_1^\ast,\,v_2^\ast,\,\dots,\,v_N^\ast)$ as
\begin{equation}
    v_i^\ast=
        \begin{cases}
            0 & \text{with probability } q \\
            a_Nv_i+b_N\displaystyle{\sum_{j\neq i}}v_i & \text{with probability } 1-q,
        \end{cases}
    \qquad i=1,\,\dots,\,N,
    \label{eq:bob1}
\end{equation}
where the constant coefficients $a_N$ and $b_N$ are expressed through the parameters $N$, $\gamma$, $s$ as
$$ a_N:=\left(1-\frac{N-1}{N}\gamma\right)s, \qquad b_N:=\frac{\gamma s}{N} \qquad (N\geq 2). $$

The economic interpretation of this toy model is the following. A group of $N$ agents produces some good to be sold in the market. The probability density $g$ models the market conditions. In particular, the two-value assumption~\eqref{eq:gt} indicates that the market is risky: there is a probability $q$ to lose all the invested capital $S$ and a probability $1-q$ to sell the produced good at a price proportional to $S$ with proportionality coefficient $s$. Thus, $s>1$ means that the trade produces a profit while $s<1$ means that part of the invested capital is lost. The two parameters $q$, $s$ in~\eqref{eq:gt} completely define the market conditions. In addition to this, the trade has also a \textit{social} part: the amount of money $S^\ast$ possibly earned from the trade is returned to the agents in accordance with certain rules tuned by the control parameter $0<\gamma<1$. In particular, the authors of~\cite{bobylev2011KRM} interpret $\gamma$ as a \textit{taxation-like} parameter, considering that a fraction $1-\gamma$ of the post-trade capital $S^\ast$ is redistributed uniformly to the traders so as to mitigate excessive differences among them.

Under some standard assumptions, the Boltzmann-type kinetic equation~\eqref{eq:kine-Gamba} resulting from the interaction rules just described may be fruitfully written in Laplace-transformed form as
\begin{equation}
    \partial_t\ff(\xi,\,t)+\ff(\xi,\,t)=Q_N(\ff)(\xi,\,t)
    \label{eq:toy}
\end{equation}
where, owing to~\eqref{eq:bob1},
$$ Q_N(\ff)(\xi,\,t)=q\ff^N(0,\,t)+(1-q)\ff(a_N\xi,\,t)\ff^{N-1}(b_N\xi,\,t), \qquad \xi\geq 0 $$
and $\ff(\xi,\,t)$ denotes the Laplace transform of $f(v,\,t)$:
$$ \ff(\xi,\,t):=\int_0^{+\infty}f(v,\,t)e^{-\xi v}\,dv. $$

As noticed in~\cite{bobylev2009CMP}, the operator $Q_N$ has a relatively simple asymptotics for large $N$, indeed
\begin{align*}
    & \ff(0,\,t)=\int_0^{+\infty}f(v,\,t)\,dv=1 \quad \text{for all } t\geq 0, \\
    & \ff(a_N\xi,\,t)\to\ff\bigl((1-\gamma)s\xi,\,t\bigr), \\
    & \ff^{N-1}(b_N\xi,\,t)=\ff^{N-1}\left(\frac{\gamma s}{N}\xi,\,t\right)\sim{\left(1-\frac{\alpha(t)\gamma s}{N}\xi\right)}^{N-1}\to e^{-\alpha(t)\gamma s\xi},
\end{align*}
where $\alpha(t):=-\partial_\xi\ff(0,\,t)$. Therefore, the asymptotic equation resulting from~\eqref{eq:toy} for $N\to\infty$ is
\begin{equation}
    \partial_t\ff(\xi,\,t)+\ff(\xi,\,t)=q+(1-q)\ff\bigl((1-\gamma)s\xi,\,t\bigr)e^{-\alpha(t)\gamma s\xi} 
    \label{eq:asy-bob}
\end{equation}
along with the boundary condition
$$ \ff(0,\,t)=1, \qquad t\geq 0. $$
Differentiating~\eqref{eq:asy-bob} with respect to $\xi$ and putting $\xi=0$ further allows one to discover
$$ \alpha(t)=e^{((1-q)s-1)t}, $$
which leads to the surprising fact that, in the limit $N\to\infty$, the Laplace-transformed Boltzmann-type kinetic equation~\eqref{eq:kine-Gamba} becomes linear and can be treated analytically.

We refer the interested reader to~\cite{bobylev2011KRM} for details on the analysis~\eqref{eq:asy-bob}, the role of the parameter $\gamma$ and the investigation of the possible formation of fat tails in the equilibrium distribution for certain values of $s$ and $q$. Here, we remark that, in spite of its simplicity, this model exhibits a great variety of interesting trends, which highlight once more the power and flexibility of the kinetic approach to the study of multi-agent systems.

\section{Online jackpot games}
\label{sect:gambling}
As documented in~\cite{wang2018PRE}, the mathematical-physical modelling of gambling activities and of their related socio-economic implications has recently gained a considerable momentum, see also \cite{Wang2019}. Typically, the goal is to understand the aggregate behaviour of a system of gamblers, aiming ultimately at adolescent gambling prevention and possibly also virtual gambling regulation. In~\cite{wang2018PRE}, the authors consider in particular the behaviour of online gamblers, that they study first by extracting a large dataset from the publicly available history page of a gambling website and then by resorting to methods of the statistical physics for the interpretation of the collected data. One of the conclusions that they draw is that the statistical distribution of the winnings exhibits, at equilibrium, a fat tail like the typical wealth curves of standard economies.

From our point of view, the huge number of gamblers and the well-defined rules of the game make it possible to treat the population of gamblers as a multi-agent economic system, in which the individuals invest part of their personal wealth in the hope to obtain a significant improvement of their economic condition. In particular, each round of the game can be modelled as a multiple interaction among a certain, possibly high, number $N$ of gamblers, who participate simultaneously in the game. In this way, resorting to kinetic equations of Boltzmann and Fokker-Planck type under suitable linearisations for $N$ large, cf.~\cite{toscani2019PRE}, we obtain a detailed interpretation of the datasets collected in~\cite{wang2018PRE}. Out of such a model-based assessment, we discover for instance that, unlike a real economy, fat tails do not actually form in the distribution of the jackpot winnings.
 
Interestingly, some related problems have been studied before. For instance, the presence of a site cut, i.e. a percentage withdrawn by the website manager from the winnings, reminds of a dissipation effect, thereby suggesting that the time evolution of the distribution function of the winnings may be described similarly to other well-known dissipative kinetic models. We recall, in particular, the model of the Maxwell-type granular gas studied in~\cite{ernst2002JSP} or the model of the Pareto tail formation in self-similar solutions of an economy undergoing recession~\cite{slanina2004PRE}. Nevertheless, unlike~\cite{ernst2002JSP,slanina2004PRE}, where the dissipation of the energy and of the mean value, respectively, was artificially restored by a suitable scaling of the variables, in this case the percentage cut on each wager is actually refilled randomly through the persistent activity of the gamblers. A second striking difference is the necessity to take into account a high number of participants in the jackpot game. In~\cite{wang2018PRE}, the authors conjecture that the shape of the steady distribution of the winnings emerging from the jackpot game does not change as the number of participants increases and, consequently, that it is sufficient to describe the evolution of the winnings for a very small number of gamblers (binary interactions in the limit). Nevertheless, as already anticipated, this does not lead to a correct interpretation of the tail of the distribution, hence of the type of economy underlying the jackpot game. Instead, by adopting a multiple-interaction kinetic description inspired by~\cite{bobylev2011KRM}, we explain that the game mechanism does not actually give rise to a power-law-type steady distribution of the winnings and therefore cannot be fully compared to a real economy.

\subsection{Maxwell-type models}
The rules of the jackpot game are quite simple: the gamblers participating in a round of the game place a bet with a certain number of lottery tickets. There is only one winning ticket in each round of the game, which is uniformly drawn among all those played in that round. The player holding the winning ticket wins all the wagers, after a site cut (percentage cut) has been subtracted.

Let us consider a number $N$ of gamblers who participate in a sequence of rounds and let $v_i\geq 0$, $i=1,\,\dots,\,N$, be the amount of money owned by the $i$th gambler at a certain point of the game. Taking into account the game rules summarised before, we express the update of the amount of money after one round of the game as
\begin{equation}
    v_i^\ast=(1-\epsilon)v_i+\epsilon(1-\delta)\sum_{j=1}^{N}v_jI(A-i)+\epsilon\beta Y_i, \qquad i=1,\,\dots,\,N,
    \label{eq:N-int.gambling}
\end{equation}
where:
\begin{enumerate*}[label=(\roman*)]
\item $0<\epsilon\leq 1$ is the fraction of money played in that round by the gamblers;
\item $0<\delta\leq 1$ is the percentage cut operated by the site on the total winning $\epsilon\sum_{j=1}^{N}v_j$;
\item $A\in\{1,\,\dots,\,N\}$ is a discrete random variable giving the index of the gambler who wins in that round;
\item $I$ is a sort of characteristic function such that $I(0)=1$ and $I(n)=0$ for all $n\in\Z\setminus\{0\}$, so that in~\eqref{eq:N-int.gambling} the gambler with $i=A$ earns all the money put into the round.
\end{enumerate*}
The further term $\epsilon\beta Y_i$ models the refilling of the amount of money available to wagers that the gamblers operate by drawing on their personal reserves of wealth. In particular, $\beta\geq 0$ is a fixed constant identifying the rate of refilling and the $Y_i$'s are non-negative, independent and identically distributed random variables giving the amount of refilled money. It is worth stressing that, in the absence of this term, the fixed cut operated by the site would cause a progressive reduction of the total amount of money in the hands of the gamblers, so that, in the long run, the gamblers would remain without money to play.

Assuming that the probability to hold the winning tickets is proportional to the number of tickets bought to play the round, we characterise the random variable $A$ via the following law:
$$ \P(A=i)=\frac{v_i}{\sum\limits_{j=1}^{N}v_j}, \qquad i=1,\,\dots,\,N. $$

Under the multiple-interaction rule~\eqref{eq:N-int.gambling}, the evolution of the distribution function $f=f(v,\,t)$ of the winnings of a prototypical gambler may be obtained by resorting to a multiple-collision Boltzmann-type model of the form~\eqref{eq:kine-Gamba}, that here we rewrite by integrating explicitly on $\R_+$:
\begin{equation}
    \frac{d}{dt}\int_{\R_+}\varphi(v)f(v,\,t)\,dv=
        \frac{1}{\tau N}\int_{\R_+^N}\sum_{i=1}^{N}\ave*{\varphi(v_i^\ast)-\varphi(v_i)}\prod_{i=1}^{N}f(v_i,\,t)\,dv_1\,\dots\,dv_N,
    \label{eq:N-Boltz.gambling}
\end{equation}
where $\ave{\cdot}$ denotes the average with respect to the distributions of the random variables $A$, $Y_i$. It is worth observing that the interaction integral on the right-hand side of~\eqref{eq:N-Boltz.gambling} has a constant unitary kernel, which, in the jargon of the classical kinetic theory, corresponds to considering \textit{Maxwellian interactions}. The reason for this choice is twofold: on one hand, it is clear that the interaction rule~\eqref{eq:N-int.gambling} guarantees $v_i^\ast\geq 0$ for all $v_i\geq 0$ and all $i=1,\,\dots,\,N$, thus all microscopic interactions~\eqref{eq:N-int.gambling} are physically admissible and none of them needs be excluded from the statistical description of the system, cf.~\cite{cordier2005JSP}. On the other hand, in the jackpot game one may realistically assume no correlation among the wagers of different gamblers, which translates perfectly in the assumption of Maxwellian interactions.

In order to better understand the role of the site cut, let us consider the evolution of the mean amount of money owned by a gambler, i.e.
$$ M_1(t):=\int_{\R_+}vf(v,\,t)\,dv. $$
Choosing $\varphi(v)=v$ in~\eqref{eq:N-Boltz.gambling} and considering that
\begin{align*}
    \ave*{\sum_{i=1}^{N}v_i^\ast} &= (1-\epsilon)\sum_{i=1}^{N}v_i+\epsilon(1-\delta)\sum_{j=1}^{N}v_j\sum_{i=1}^{N}\P(A=i)
        +\epsilon\beta\sum_{i=1}^{N}\ave{Y_i} \\
    &= (1-\epsilon\delta)\sum_{i=1}^{N}v_i+N\epsilon\beta m,
\end{align*}
where we have denoted by $m\geq 0$ the mean of the $Y_i$'s, we discover
\begin{equation}
    \frac{dM_1}{dt}=-\frac{\epsilon\delta}{\tau}M_1+\frac{\epsilon\beta}{\tau}m.
    \label{eq:M1.gambling}
\end{equation}
As expected, the presence of a percentage cut $\delta>0$ operated by the website manager leads to an exponential decay of the mean at a rate proportional to $\delta$ itself, which however may be compensated, on average, by the refilling operated by the players to continue to gamble.
 
As far as the computation of higher order moments of $f$ is concerned, analytic results may instead not always be obtained due to the strong non-linearity of the Boltzmann-type equation~\eqref{eq:N-Boltz.gambling}.

\subsection{Linearised model for large~\texorpdfstring{$\boldsymbol{N}$}{}}
Although it gives a precise picture of the evolution of the jackpot game, the highly non-linear Boltzmann-type equation~\eqref{eq:N-Boltz.gambling} has essentially to be treated numerically in order to extract from it useful detailed information.

Nevertheless, a considerable simplification occurs for a large number $N$ of gamblers. Indeed, in this situation we have
\begin{equation}
    \sum_{i=1}^{N}v_i=N\cdot\frac{1}{N}\sum_{i=1}^{N}v_i\approx NM_1(t),
    \label{eq:linearisation.gambling}
\end{equation}
which corresponds to approximating the empirical mean wealth of the gamblers participating in a round with the theoretical mean wealth owned by the entire population of potential gamblers. As a consequence, the interaction rule~\eqref{eq:N-int.gambling} may be linearised as
\begin{equation}
    v^\ast=(1-\epsilon)v+N\epsilon(1-\delta)M_1(t)I(\bar{A}-1)+\epsilon\beta Y,
    \label{eq:lin-int.gambling}
\end{equation}
where we have suppressed the index $i$ of the gamblers as it is now inessential. The new discrete random variable $\bar{A}\in\{0,\,1\}$ indicates simply whether the generic gambler wins ($\bar{A}=1$) or not ($\bar{A}=0$) in a single round. In view of the linearisation~\eqref{eq:linearisation.gambling}, we deduce the law of $\bar{A}$ from that of $A$ as
$$ \P(\bar{A}=1)=\frac{v}{NM_1(t)}, \qquad \P(\bar{A}=0)=1-\frac{v}{NM_1(t)}. $$
In particular, we notice that the usual properties $0\leq\P(\bar{A}=0),\,\P(\bar{A}=1)\leq 1$ might not be strictly satisfied if $N$ is not large enough, but are more and more met as $N$ grows. This is actually not a major problem, because we will be mostly interested in the asymptotic regime $N\to \infty$.

As a consequence of the new interaction rule~\eqref{eq:lin-int.gambling}, the Boltzmann-type equation describing the evolution of the distribution function $f$ of the gambler's winnings linearises as well as
\begin{equation}
    \frac{d}{dt}\int_{\R_+}\varphi(v)f(x,\,t)\,dv=\frac{1}{\tau}\int_{\R_+}\ave{\varphi(v^\ast)-\varphi(v)}f(v,\,t)\,dv.
    \label{eq:lin-Boltz.gambling}
\end{equation}
This equation makes possible an explicit computation of the statistical moments of $f$. In particular, by choosing $\varphi(v)=v$, it is easy to see that it provides the same evolution of the first moment as~\eqref{eq:M1.gambling}.

Nevertheless, we point out that if the fraction $\epsilon$ of money played by each gambler is fixed independently of the total number $N$ of gamblers, the total fraction $\epsilon N$ of money played in a single round tends to blow as $N$ increases, which is not consistent with the jackpot game. Therefore, we further assume that the product $\epsilon N$ actually equals a constant value, say $\kappa>0$, so that the percent amount of money played in each round remains finite for every $\epsilon$, $N$:
\begin{equation}
    \epsilon N=:\kappa.
    \label{eq:kappa.gambling}
\end{equation}

\subsection{Steady distributions with slim tails}
We now take advantage of the linearised model~\eqref{eq:lin-int.gambling},~\eqref{eq:lin-Boltz.gambling} to investigate the shape of the large time statistical distribution of the gambler's winnings. To this purpose, we rely on the asymptotic procedure of the quasi-invariant limit, upon observing from~\eqref{eq:lin-int.gambling} that in the regime of small $\epsilon$, or equivalently of large $N$ owing to~\eqref{eq:kappa.gambling}, interactions are indeed quasi-invariant. In order to approach the steady state in such a regime, we scale the relaxation time scale as $\tau=\epsilon$ in~\eqref{eq:lin-Boltz.gambling} and we choose $\varphi(v)=e^{-i\xi v}$, where now $i$ denotes the imaginary unit and $\xi\in\R$. We obtain therefore the following time-scaled Fourier-transformed version of the kinetic equation~\eqref{eq:lin-Boltz.gambling}:
$$ \partial_t\ff(\xi,\,t)=\frac{1}{\epsilon}\int_{\R_+}\ave*{e^{-i\xi v^\ast}-e^{-i\xi v}}f(v,\,t)\,dv, $$
where $\ff$ denotes the Fourier transform of the distribution function $f$:
$$ \ff(\xi,\,t):=\int_{\R_+}f(v,\,t)e^{-i\xi v}\,dv. $$
Taking now the limit $\epsilon\searrow 0$, see~\cite{toscani2019PRE} for the details, this procedure shows that for a large number $N$ of gamblers and on a large time scale the non-linear kinetic model~\eqref{eq:N-int.gambling},~\eqref{eq:N-Boltz.gambling} is well approximated by the Fourier-transformed linear equation
\begin{equation}
    \partial_t\ff=\left[\frac{i}{\kappa M_1(t)}\left(e^{-i\kappa M_1(t)(1-\delta)\xi}-1\right)-\xi\right]\partial_\xi\ff-i\beta m\xi\ff.
    \label{eq:Fourier.gambling}
\end{equation}

In order to gain further insights into the physical variable $v$, it is useful to consider in particular the regime of small $\kappa$. Then
$$ \left[\frac{i}{\kappa M_1(t)}\left(e^{-i\kappa M_1(t)(1-\delta)\xi}-1\right)-\xi\right]\partial_\xi\ff\approx
    \left[-\delta\xi-\frac{i\kappa M_1(t)}{2}(1-\delta)^2\xi^2\right]\partial_\xi\ff $$
and, within this approximation, we transform back~\eqref{eq:Fourier.gambling} as
\begin{equation}
    \partial_tf=\frac{\kappa(1-\delta)^2M_1(t)}{2}\partial_v^2(vf)+\partial_v\bigl((\delta v-\beta m)f\bigr),
    \label{eq:FP.gambling}
\end{equation}
which is a Fokker-Planck equation with time-dependent diffusion coefficient. Recalling from~\eqref{eq:M1.gambling} that, in the time scaling $t\to t/\epsilon$, it results $M_1(t)\to\frac{\beta}{\delta}m$ for $t\to +\infty$, we are in a position to identify from~\eqref{eq:FP.gambling} the unique stationary solution with unitary mass, say $f^\infty$, which plays here the role of the Maxwellian distribution in the classical kinetic theory:
\begin{equation}\label{eq:gamma}
 f^\infty(v)=\frac{\left(\frac{\mu}{\beta m}\right)^\mu}{\Gamma(\mu)}v^{\mu-1}e^{-\frac{\mu}{\beta m}v},
    \qquad \mu:=\frac{2\delta^2}{\kappa(1-\delta)^2}. 
\end{equation}
Such an $f^\infty$ is a gamma probability density function, which has clearly bounded moments of any order. Therefore, we conclude that no fat tail is produced in this case in the stationary distribution of the gambler's winnings.

As a matter of fact, this results proves the uniform boundedness of all moments of $f^\infty$, hence the slimness of its tail, only for the linearised kinetic model~\eqref{eq:lin-int.gambling},~\eqref{eq:lin-Boltz.gambling} in the limit regime $\epsilon\searrow 0$ or, equivalently, $N\to\infty$. Nevertheless, the theory developed so far suggests that also the ``real'' kinetic model~\eqref{eq:N-int.gambling},~\eqref{eq:N-Boltz.gambling} may behave in the same way, as it will be indeed confirmed by the numerical tests in Section~\ref{sect:numerics}.

\section{Taxation and wealth redistribution}
\label{sect:taxation}
Kinetic models of wealth distribution in a multi-agent society often tried to take into account the effects of realistic features, such as taxation and redistribution. One of the the first attempts to deal with the problem of taxation by means of simple stochastic models may be found in~\cite{garibaldi2007EPJB}. There, a simple stochastic exchange game mimicking taxation and redistribution is introduced and its large-time trend studied in detail. Specifically, the taxation mechanism is modelled by extracting randomly some agents, whose wealth is redistributed to other agents according to the P\'{o}lya's urn scheme. In the continuum limit, the individual wealth  distribution is shown to converge to a gamma probability density, whose form factor coincides with the redistribution weight.

A different attempt may be found in~\cite{guala2009INDECS}, where it is suggested that inelastic binary collisions may be regarded as the application of taxes and that their redistribution may reproduce the salient features of empirical wealth distributions. The model in~\cite{guala2009INDECS} is reminiscent of the inelastic kinetic model introduced in~\cite{slanina2004PRE}. It takes into account a simple granular closed-system model, in which the collisions are inelastic and the loss of energy is redistributed among the particles of the system according to a certain criterion.

Parellelly to the contributions just mentioned, classical kinetic models of wealth redistribution have been formulated also at the continuous level, taking advantage of kinetic Boltzmann-type equations. In this case, either binary interactions~\cite{bisi2009CMS} or interactions with a background~\cite{toscani2009EPL} are dominant. In particular, in~\cite{bisi2009CMS,toscani2009EPL} the novelty was to introduce a simple taxation mechanism at the level of the single trade, which produces a portion of wealth subsequently redistributed to the agents according to some precise rules. The redistribution mechanism is assumed to be sufficiently flexible to return to the agents either a constant amount of wealth, independent of the agent's wealth itself, or an amount of wealth proportional (or inversely proportional) to the agent's wealth. In these models, the redistribution mechanism is conceived in such a way to keep the mean wealth of the system constant in time. Such a conservation allows the statistical distribution of the system to reach a certain asymptotic profile, which may provide information on the effect of the underlying taxation mechanism~\cite{pareschi2013BOOK,toscani2009EPL}.

Taxation and consequent wealth redistribution furnish a further prototypical example of multiple interactions in a multi-agent society, that can be treated by means of the methodology introduced in~\cite{bobylev2009CMP}. In particular, the forthcoming analysis provides physical bases to recognise whether the redistribution mechanism described in~\cite{bisi2009CMS,toscani2009EPL} directly at the level of \textit{binary} interactions is actually consistent with a more realistic \textit{multiple} interaction setting.

Let us consider a population of $N$ agents and let $v_i\geq 0$ be the wealth of the $i$th agent. We suppose that each agent performs economical transactions with an external background, whose wealth is modelled by a microscopic variable $w\geq 0$ sampled from a known distribution with given probability density $g=g(w)$. For the sake of simplicity, we assume that the background distribution does not change in time. We describe the update of the wealth $v_i$ after a transaction as:
\begin{equation}
    v_i^\ast=(1-\lambda)v_i+\lambda w+v_i\eta_i-\alpha(\lambda+\eta_i)v_i+\frac{\alpha}{N}\sum_{j=1}^{N}(\lambda+\eta_j)v_j,
        \qquad i=1,\,\dots,\,N
    \label{eq:N-int.taxation}
\end{equation}
where $\lambda\in (0,\,1]$ is the percent amount of wealth that the $i$th agent invests in the transaction, $\eta_i\in\R$ is a random variable modelling the intrinsic risk of the transaction and $\alpha\in [0,\,1]$ is the rate of taxation. We assume that the $\eta_i$'s are independent and identically distributed, in particular with zero mean and strictly positive variance:
\begin{equation}
    \ave{\eta_i}=0, \qquad \ave{\eta_i^2}=:\sigma^2>0,
    \label{eq:eta_i}
\end{equation}
As before, $\ave{\cdot}$ denotes the average with respect to the common distribution of the $\eta_i$'s. In essence, in~\eqref{eq:N-int.taxation} the term $-\alpha(\lambda+\eta_i)v_i$ establishes that a percentage $\alpha$ of the invested wealth is taxed, taking duly into account the actual risk of the investment; while the term $\frac{\alpha}{N}\sum_{j=1}^{N}(\lambda+\eta_j)v_j$ means that the total wealth obtained from the taxation is evenly redistributed to all the individuals of the society (for instance, in terms of supplied services).

\subsection{Analysis of the microscopic interaction rule}
\label{sect:int_analysis.taxation}
In order to be physically admissible, rule~\eqref{eq:N-int.taxation} has to guarantee that $v_i^\ast\geq 0$, $i=1,\,\dots,\,N$, for all $v_1,\,\dots,\,v_N,\,w\geq 0$. A sufficient condition for this is clearly
$$ (1-\lambda+\eta_i-\alpha\lambda-\alpha\eta_i)v_i+\frac{\alpha}{N}\sum_{j=1}^{N}(\lambda+\eta_j)v_j\geq 0, $$
which is certainly satisfied if e.g.,
$$  \begin{cases}
        1-(1+\alpha)\lambda+(1-\alpha)\eta_i\geq 0 \\
        \eta_i\geq -\lambda,
    \end{cases}
    \qquad \forall\,i=1,\,\dots,\,N $$
and finally if
$$ \eta_i\geq\max\left\{-\lambda,\,-\frac{1-(1+\alpha)\lambda}{1-\alpha}\right\} \qquad \forall\,i=1,\,\dots,\,N. $$
Moreover, in order to guarantee the fulfillment of~\eqref{eq:eta_i} it is necessary that $1-(1+\alpha)\lambda>0$, i.e. $\lambda<\frac{1}{1+\alpha}$. In conclusion, under these conditions a truncation from the left of the support of the $\eta_i$'s provides the necessary physical consistency to the interaction rule~\eqref{eq:N-int.taxation}.

\subsection{Boltzmann-type kinetic description}
In view of the analysis of the microscopic interactions just performed, we conclude that a Maxwellian kinetic model, i.e. one with constant interaction kernel, is appropriate for the particle system at hand. Indeed, under the restrictions on the $\eta_i$'s set forth above, none of the microscopic interactions~\eqref{eq:N-int.taxation} needs to be excluded from the statistical description of the system, cf.~\cite{cordier2005JSP,toscani2006CMS}. Therefore, if $f=f(v,\,t)$ denotes the probability density function of the wealth of a generic individual at time $t>0$, the Boltzmann-type equation ruling the evolution of $f$ under the multiple interaction rule~\eqref{eq:N-int.taxation} is of the form~\eqref{eq:kine-Gamba} and, in particular, for this application reads
\begin{equation}
    \frac{d}{dt}\int_{\R_+}\varphi(v)f(v,\,t)\,dv=\frac{1}{\tau N}\int_{\R_+^{N+1}}\sum_{i=1}^{N}\ave{\varphi(v_i^\ast)-\varphi(v_i)}\prod_{i=1}^{N}f(v_i,\,t)g(w)\,dv_1\,\dots\,dv_N\,dw.
    \label{eq:N-Boltz.taxation}
\end{equation}
Choosing $\varphi(v)=v$ and denoting by
$$ M_1(t):=\int_{\R_+}vf(v,\,t)\,dv, \qquad m:=\int_{\R_+}wg(w)\,dw $$
the mean wealth of the agents and of the background, respectively, by a direct calculation we find that
\begin{equation}
    \frac{dM_1}{dt}=\frac{\lambda}{\tau}(m-M_1),
    \label{eq:M1.taxation}
\end{equation}
namely that $M_1=M_1(t)$ relaxes exponentially fast in time on $m$.

By far more intricate is to obtain from~\eqref{eq:N-int.taxation},~\eqref{eq:N-Boltz.taxation} the evolution of the variance of the wealth distribution, which nevertheless provides useful information on the effectiveness of the taxation-redistribution policy in mitigating social inequalities. In order to investigate this issue, a linearisation of the multiple-interaction model for $N$ large is a particularly fruitful strategy.

\subsection{Linearised model for large~\texorpdfstring{$\boldsymbol{N}$}{}}
In~\eqref{eq:N-int.taxation}, the term responsible for the multiplicity of simultaneous interactions is clearly the redistribution one:
$$ R_N:=\frac{\alpha}{N}\sum_{j=1}^{N}(\lambda+\eta_j)v_j
    =\frac{\alpha\lambda}{N}\sum_{j=1}^{N}v_j+\frac{\alpha}{N}\sum_{j=1}^{N}v_j\eta_j. $$
Considering that the $v_j$'s are sampled from the probability distribution described by $f$, when $N$ is sufficiently large we have $\frac{1}{N}\sum_{j=1}^{N}v_j\approx M_1$. Concerning the other term contributing to $R_N$, we observe that
$$ \ave*{\frac{\alpha}{N}\sum_{j=1}^{N}v_j\eta_j}=0, \qquad
    \Var\left(\frac{\alpha}{N}\sum_{j=1}^{N}v_j\eta_j\right)=\frac{\alpha^2\sigma^2}{N^2}\sum_{j=1}^{N}v_j^2. $$
If we assume that the second moment of $f$ is bounded, so that $\frac{1}{N}\sum_{j=1}^{N}v_j^2$ remains bounded for all $N$, we deduce that $\Var\left(\frac{\alpha}{N}\sum_{j=1}^{N}v_j\eta_j\right)\to 0$ as $N\to\infty$, which then implies that the random variable $\frac{\alpha}{N}\sum_{j=1}^{N}v_j\eta_j$ converges to zero almost surely. Therefore, for $N$ sufficiently large we may well approximate $R_N$ as
$$ R_N\approx\alpha\lambda M_1 $$
and consider the linearised time-dependent interaction rule
\begin{equation}
    v^\ast=(1-\lambda)v+\lambda w+v\eta-\alpha(\lambda+\eta)v+\alpha\lambda M_1(t),
    \label{eq:lin-int.taxation}
\end{equation}
where we have dropped the index $i$, which is now inessential. Arguing like before, we see that a sufficient condition guaranteeing that $v^\ast\geq 0$ for all $v,\,w\geq 0$ is
$$ \eta\geq -\frac{1-(1+\alpha)\lambda}{1-\alpha} $$
together with $\lambda<\frac{1}{1+\alpha}$, which is indeed consistent with the previous findings.

In view of this, a Maxwellian kinetic model is appropriate also in this case. In particular, the Boltzmann-type kinetic model reads now
\begin{equation}
    \frac{d}{dt}\int_{\R_+}\varphi(v)f(v,\,t)\,dv=\frac{1}{\tau}\int_{\R_+^2}\ave{\varphi(v^\ast)-\varphi(v)}f(v,\,t)g(w)\,dv\,dw
    \label{eq:lin-Boltz.taxation}
\end{equation}
whence, by letting $\varphi(v)=v$, we discover that the time evolution of the mean wealth of the population is again ruled by~\eqref{eq:M1.taxation}. Hence, the linearised model preserves the average trend of the full multiple-interaction model.

To characterise the large time trend of the variance $\Sigma$ of the wealth distribution:
$$ \Sigma(t):=M_2(t)-M_1^2(t) \qquad \text{with} \qquad M_2(t):=\int_{\R_+}v^2f(v,\,t)\,dv, $$
it is convenient to consider the quasi-invariant interaction regime: one assumes to be sufficiently close to equilibrium, so that each interaction produces a very small variation of wealth from $v$ to $v^\ast$. Upon introducing a small parameter $\epsilon>0$, this regime may be mimicked by setting $\lambda=\sigma^2=\epsilon$ in~\eqref{eq:lin-int.taxation} (small effect of the interactions) and also $\tau=\epsilon$ in~\eqref{eq:lin-Boltz.taxation} (larger time scale, closeness to equilibrium). Notice that the aforesaid scaling of the variance of $\eta$ amounts to writing $\eta=\sqrt{\epsilon}Y$, where $Y$ is the standardisation of $\eta$ (thus, in particular, $\ave{Y}=0$ and $\ave{Y^2}=1$). This leads to the scaled interaction rule
$$ v^\ast=(1-\epsilon)v+\epsilon w+\sqrt{\epsilon}vY-\alpha(\epsilon+\sqrt{\epsilon}Y)v+\alpha\epsilon M_1(t) $$
and to the scaled Boltzmann-type equation
\begin{equation}
    \frac{d}{dt}\int_{\R_+}\varphi(v)f(v,\,t)\,dv=\frac{1}{\epsilon}\int_{\R_+^2}\ave{\varphi(v^\ast)-\varphi(v)}f(v,\,t)g(w)\,dv\,dw.
    \label{eq:lin-Boltz_eps.taxation}
\end{equation}

For $\varphi(v)=v$ we have
\begin{equation}
    \frac{dM_1}{dt}=m-M_1,
    \label{eq:quasi-inv.M1}
\end{equation}
while for $\varphi(v)=v^2$ and in the asymptotic regime $\epsilon\searrow 0$ we obtain
\begin{equation}
    \frac{dM_2}{dt}=\left((2-\alpha)^2-5\right)M_2+2(m+\alpha M_1)M_1.
    \label{eq:quasi-inv.M2}
\end{equation}
In particular,~\eqref{eq:quasi-inv.M1} gives the evolution of the mean wealth for every $\epsilon>0$, whereas~\eqref{eq:quasi-inv.M2} approximates the large time trend of the second moment of $f$ for any sufficiently small value of $\epsilon$.

From~\eqref{eq:quasi-inv.M1} we see that $M_1\to M_1^\infty:=m$ as $t\to+\infty$. Consequently, upon observing that $(2-\alpha)^2-5<0$ for all $\alpha\in [0,\,1]$, from~\eqref{eq:quasi-inv.M2} we deduce that
$$ M_2\to M_2^\infty:=\frac{2(1+\alpha)}{5-(2-\alpha)^2}m^2 $$
when $t\to+\infty$. Thus the limit value, say $\Sigma^\infty$, approached by the variance of the wealth distribution for large times is
\begin{equation}
    \Sigma^\infty=M_2^\infty-\left(M_1^\infty\right)^2=\frac{(1-\alpha)^2}{5-(2-\alpha)^2}m^2.
    \label{eq:Sigmainf}
\end{equation}

\begin{figure}[t]
\centering
\includegraphics[scale=0.6]{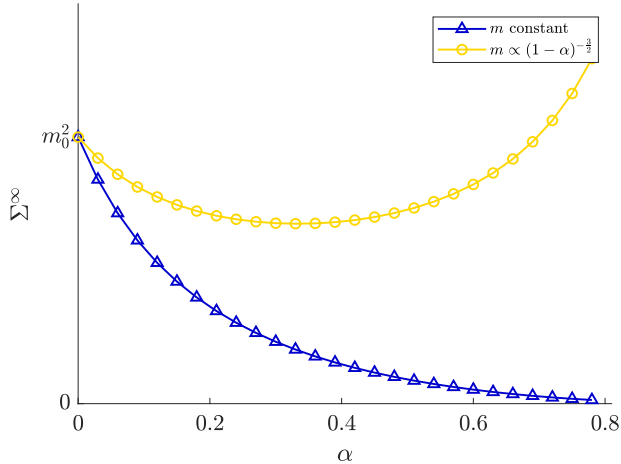}
\caption{The asymptotic wealth variance $\Sigma^\infty$ vs. the taxation rate $\alpha$ for both constant and $\alpha$-dependent mean wealth $m$ of the background.}
\label{fig:Sigmainf}
\end{figure}

Meaningful considerations may be made by inspecting the trend of $\Sigma^\infty$ with respect to the taxation rate $\alpha$. If the mean wealth $m$ of the background is constant then $\Sigma^\infty$ decreases monotonically to zero with $\alpha$, meaning that the maximum of social equality is obtained with a full taxation of the invested capital, see Figure~\ref{fig:Sigmainf}. A more interesting scenario is obtained if we assume that $m$ varies instead with $\alpha$ and, in particular, that it increases as $\alpha$ approaches $1$. This models the fact that the background benefits from the taxation of the population. Following this idea, we may set
$$ m=\frac{m_0}{(1-\alpha)^\gamma}, \qquad m_0,\,\gamma>0, $$
so that the limit value of the variance of the wealth distribution becomes
$$ \Sigma^\infty=\frac{(1-\alpha)^{2(1-\gamma)}}{5-(2-\alpha)^2}m_0^2. $$
A standard computation shows that, depending on $\gamma$, this function may reach a global minimum for $\alpha\in (0,\,1)$, meaning that there exists an optimal taxation rate which guarantees the least social inequality. For instance, for $\gamma=\frac{3}{2}$ such a minimum is attained for $\alpha=\frac{1}{3}\approx 33\%$, see Figure~\ref{fig:Sigmainf} again. In spite of the fact that the society becomes richer and richer on average for $\alpha\nearrow 1$ (because $M_1$ relaxes to $m$ as $t\to+\infty$ and $m$ blows as $\alpha\nearrow 1$), we notice that in this case the optimal taxation rate is realistically not $\alpha=1$. The reason is that now also the variance blows as $\alpha\nearrow 1$, thus the richness accumulated on average by the society tends to be unevenly distributed for too high taxation rates.

\begin{remark}
Let us momentarily go back to the multiple-interaction rule~\eqref{eq:N-int.taxation} and let us consider the quasi-invariant regime also in~\eqref{eq:N-Boltz.taxation}. Thus we set $\lambda=\tau=\epsilon$, $\eta_i=\sqrt{\epsilon}Y_i$ with $\ave{Y_i}=0$ and $\ave{Y_i^2}=1$ for all $i=1,\,\dots,\,N$. For $\varphi(v)=v^2$, some technical calculations show that, in the asymptotic limit $\epsilon\searrow 0$, it holds
\begin{equation}\label{eq:tax_energy}
 \frac{dM_2}{dt}=\left((2-\alpha)^2-5\right)M_2+2(m+\alpha M_1)M_1-\frac{\alpha^2}{N}M_2. 
 \end{equation}
This equation differs from~\eqref{eq:quasi-inv.M2} only in the last term on the right-hand side, which however vanishes for $N\to\infty$. Consequently, we conclude that the second order moment, hence also the wealth variance, produced by the linearised model is consistent with that of the original multiple-interaction model in the limit of a high number of individuals participating in the simultaneous interactions.
\end{remark}

By applying the quasi-invariant limit procedure to~\eqref{eq:lin-Boltz_eps.taxation} with a sufficiently smooth and compactly supported test function $\varphi$, for instance $\varphi\in C^3_c(\R_+)$, we may further detail the previous picture by determining an explicit asymptotic approximation of the whole distribution function $f$ valid for large times in the regime of small interaction parameters. In particular, for $\epsilon\searrow 0$ we find that $f$ satisfies the Fokker-Planck equation
\begin{equation}
    \partial_tf=\frac{(1-\alpha)^2}{2}\partial^2_v(v^2f)-\partial_v[(m+\alpha M_1(t)-(1+\alpha)v)f],
    \label{eq:FP.taxation}
\end{equation}
which, recalling that $M_1\to m$ for $t\to +\infty$, if $\alpha<1$ admits the following unique stationary solution with unitary mass (the analogous of the Maxwellian in the classical kinetic theory):
\begin{equation}
    f^\infty(v)=\frac{(\mu m)^{1+\mu}}{\Gamma(1+\mu)}\cdot\frac{e^{-\frac{\mu m}{v}}}{v^{2+\mu}},
        \qquad \mu:=\frac{2(1+\alpha)}{(1-\alpha)^2}.
    \label{eq:inv_gamma}
\end{equation}

Such an $f^\infty$ is an inverse gamma probability density function with mean $m$ and variance $\Sigma^\infty$, see~\eqref{eq:Sigmainf}. Consistently with standard wealth distribution curves, it exhibits a Pareto-type fat tail~\cite{gualandi2018ECONOMICS} with Pareto exponent (or index) equal to $1+\mu$. We observe that the Pareto exponent depends explicitly on the taxation rate $\alpha$ and, in particular, that it increases as $\alpha\nearrow 1$. Such an exponent is the same for both a constant and an $\alpha$-dependent mean wealth $m$ of the background. In particular, in both cases high taxation rates make the tail of the distribution slimmer and slimmer.

For $\alpha=1$, the expression~\eqref{eq:inv_gamma} of $f^\infty$ ceases to be valid. Going back to~\eqref{eq:FP.taxation}, one realises that in such a case the stationary distribution becomes $f^\infty(v)=\delta(v-m)$, which clearly makes sense only if $m$ does not blow for $\alpha\nearrow 1$.

\begin{remark}
Fokker-Planck equations like~\eqref{eq:FP.taxation} modelling wealth redistribution in presence of taxation have been recently considered also in~\cite{bisi2017BUMI}, where the quasi-invariant limit is applied to the Boltzmann-type equation with taxation proposed in~\cite{bisi2009CMS}. In agreement with the present case, the effect of taxation at the level of the kinetic equation is to increase the value of the Pareto index featured by the steady state of the Fokker-Planck equation. Unlike~\eqref{eq:FP.taxation}, however, the drift term in the Fokker-Planck equation derived in~\cite{bisi2017BUMI} does not depend on time.

At the level of binary interactions, taxation effects may be replaced by a control aiming to minimise the Gini coefficient~\cite{duering2018EPJB}. Also in this case, the Fokker-Planck equation resulting in the quasi-invariant regime coincides with the one considered in~\cite{bisi2017BUMI} and produces an equilibrium distribution with a higher Pareto index than in the uncontrolled case.
\end{remark}

\section{Numerical tests}
\label{sect:numerics}
In this section, we provide numerical insights into the various models discussed before, resorting to direct Monte Carlo methods for collisional kinetic equations and to the recent structure preserving methods for Fokker-Planck equations. For a comprehensive presentation of numerical methods for kinetic equations, we refer the interested reader to~\cite{dimarco2014AN,pareschi2013BOOK,pareschi2018JSC}.

As a first step towards a more detailed study of numerical methods for Maxwellian kinetic equations with multiple interactions, we consider classical direct simulation Monte Carlo methods.  We rewrite~\eqref{eq:kine-Gamba} in strong form as
\begin{align}
    \begin{aligned}[b]
        \partial_t f(v,\,t) &= \frac{1}{\tau}\ave*{\int_{\R_+^{N-1}}\left(\frac{1}{\pr{J}}\prod_{i=1}^Nf(\pr{v}_i,\,t)-\prod_{i=1}^Nf(v_i,\,t)\right)\,dv_2\,\dots\,dv_N} \\
        &= \frac{1}{\tau}\left(Q^+(f,\,\dots,\,f)(v,\,t)-f(v,\,t)\right),
    \end{aligned}
    \label{eq:kine1_strong}
\end{align}
where $Q^+$ is the gain operator:
$$ Q^+(f,\,\dots,\,f)(v,\,t):=\left\langle \int_{\R_+^{N-1}}\prod_{i=1}^{N}\frac{1}{\pr{J}}f(\pr{v}_i,\,t)\,dv_2\,\dots\,dv_N \right\rangle $$
and $\pr{J}$ is the Jacobian of the transformation from the pre-interaction variables $\{\pr{v}_i\}_{i=1}^{N}$ to the post-interaction variables $\{v_i\}_{i=1}^{N}$. Furthermore, $N>1$ is the number of agents simultaneously interacting in a single interaction and $\ave{\cdot}$ denotes the expectation taken with respect to the random part of the microscopic interactions.

We introduce a time mesh $t^n:=n\Delta{t}$, $n\ge 1$, $\Delta{t}>0$, and we consider the following time-discrete version of~\eqref{eq:kine1_strong} through a forward scheme:
$$ f^{n+1}(v)=\left(1-\frac{\Delta{t}}{\tau}\right)f^n(v)+\frac{\Delta{t}}{\tau}Q^+(f^n,\,\dots,\,f^n)(v), $$
where $f^n(v):=f(v,\,t^n)$. We observe that by choosing $\Delta{t}=\tau$ the first term on the right-hand side (loss part of the equation) disappears, hence at each time step only the gain operator $Q^+$ needs to be computed.

In the following, we provide numerical evidence of the consistency of the linearisation of the multiple-interaction Boltzmann-type models discussed in the previous sections. More specifically, we show that for $N\gg 1$ the dynamics described by those models are indeed nicely caught by the corresponding linear models with suitably modified interactions and that the latter are also able to capture the stationary distributions in the quasi-invariant interaction regime. We stress that this represents a real reduction of complexity with respect to the original multiply collisional kinetic models: indeed, it makes it possible to resort to reduced models, such as linear kinetic equations and Fokker-Planck-type operators, for which detailed analytical and numerical insights are accessible.

\subsection{Online jackpot games}
In Section~\ref{sect:gambling} we presented a kinetic model with multiple interactions for virtual-item jackpot games. We recall that the microscopic dynamics are given by~\eqref{eq:N-int.gambling}. To model the background distribution, we rely on the detailed discussion in~\cite{toscani2019PRE}, which is essentially based on the results of~\cite{dimarco2019JSP,gualandi2019M3AS}. In particular, we consider for the $Y_i$'s a log-normal probability density with unitary mean $m=\ave{Y_i}=1$, specifically:
$$ Y_i\sim\frac{1}{\sqrt{4\pi}y}\exp\left(-\frac{{(\log{y}+1)}^2}{2}\right). $$
Furthermore, in all the numerical tests we fix $\tau=0.1$ and $\beta=\delta=0.2$, meaning that the rate of refilling by the gamblers equals the percentage of the site cut. See~\cite{toscani2019PRE} for a systematic study of other relevant cases. 

We solve both the multiple-interaction and the linearised Boltzmann-type equations~\eqref{eq:N-Boltz.gambling},~\eqref{eq:lin-Boltz.gambling} by a direct simulation Monte Carlo scheme, following the ideas summarised at the beginning of Section~\ref{sect:numerics}. We consider a random sample of $10^6$ particles with initial uniform distribution of the winnings for $v\in [0,\,2]$, thus
\begin{equation}
    f_0(v)=f(v,\,0)=\frac{1}{2}\chi_{[0,\,2]}(v),
    \label{eq:f0}
\end{equation}
where $\chi_A$ is the characteristic function of the set $A$.

\begin{figure}[!t]
\centering
\subfigure[$t=1$]{\includegraphics[scale=0.38]{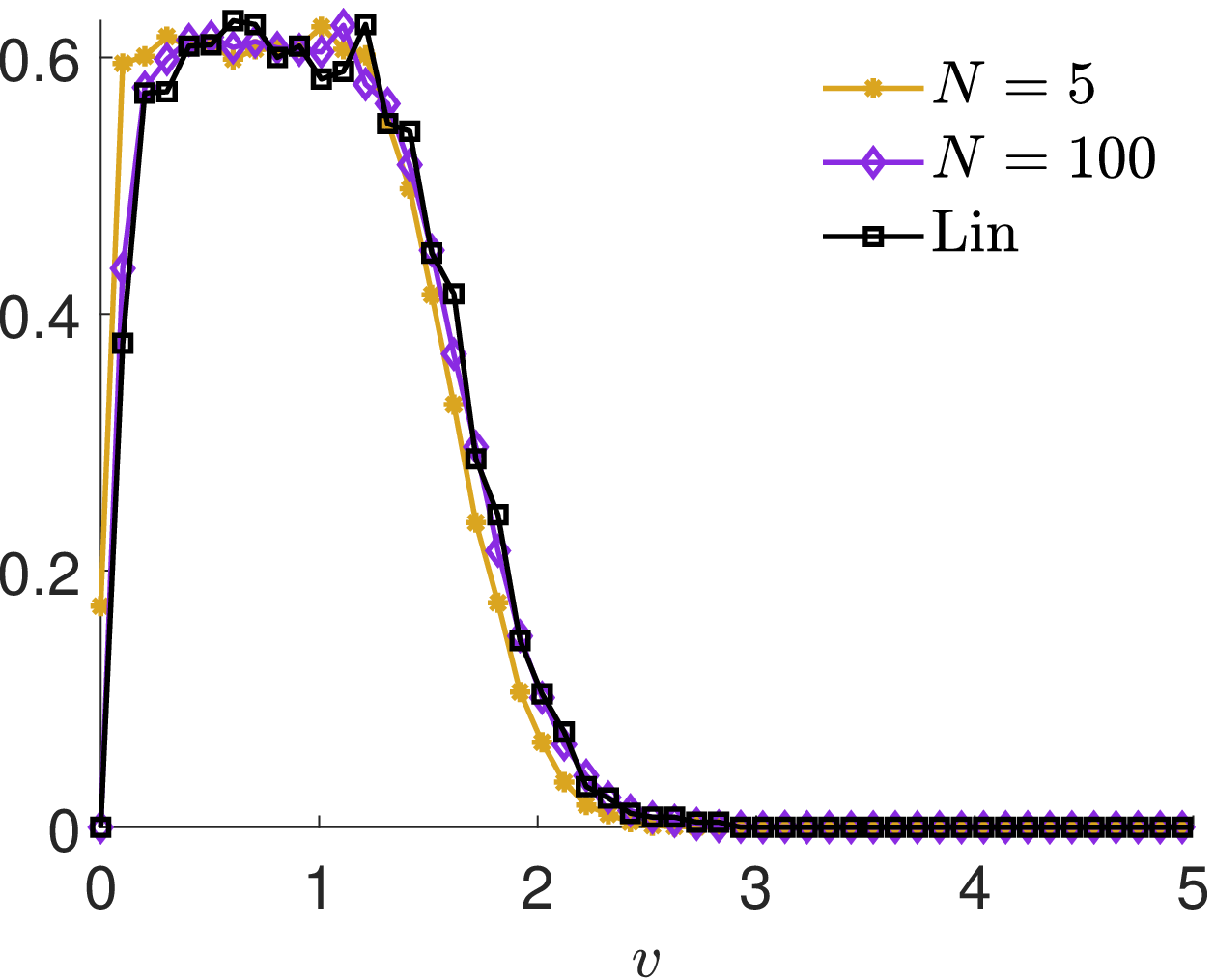}}
\subfigure[$t=5$]{\includegraphics[scale=0.38]{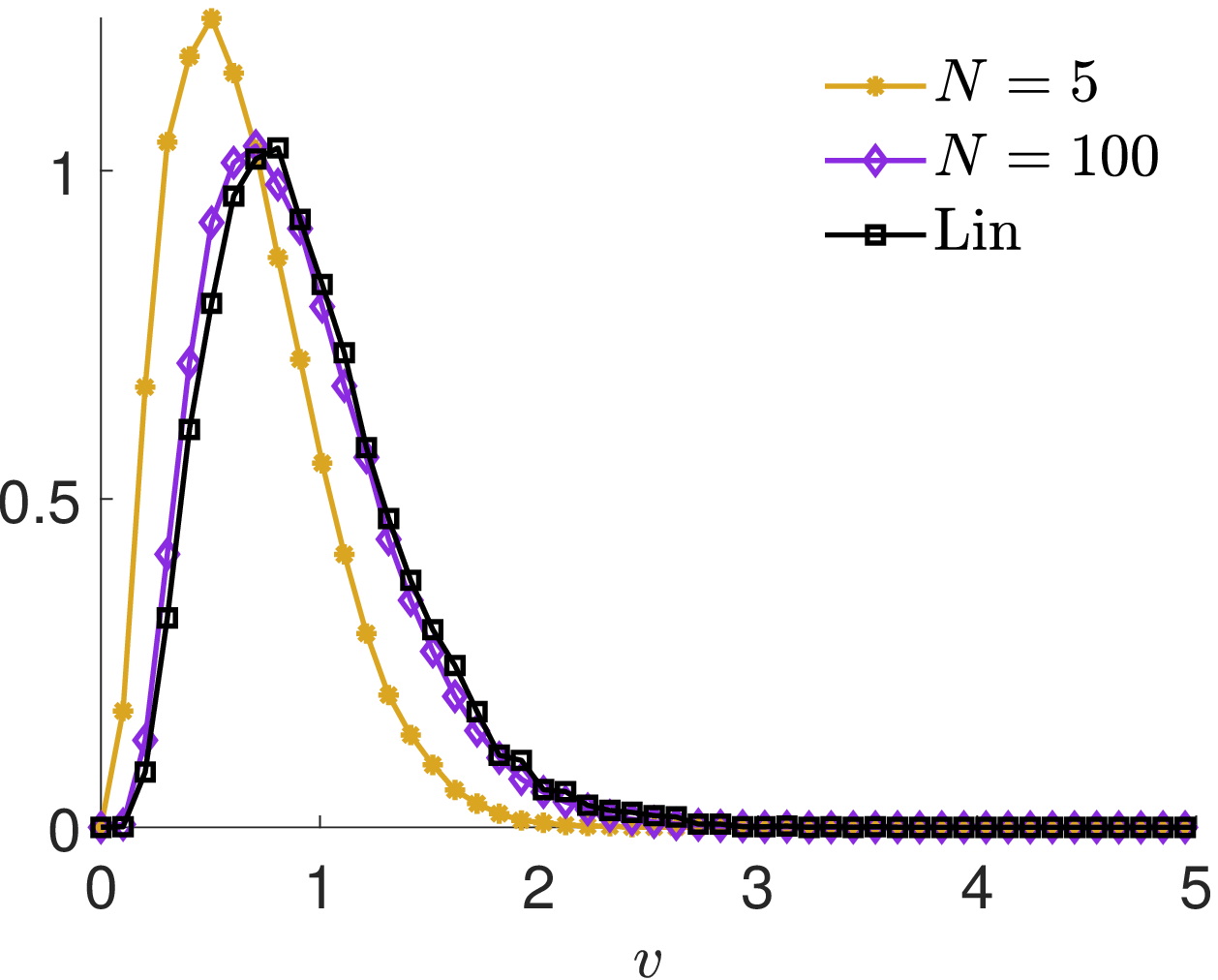}}
\subfigure[$t=25$]{\includegraphics[scale=0.38]{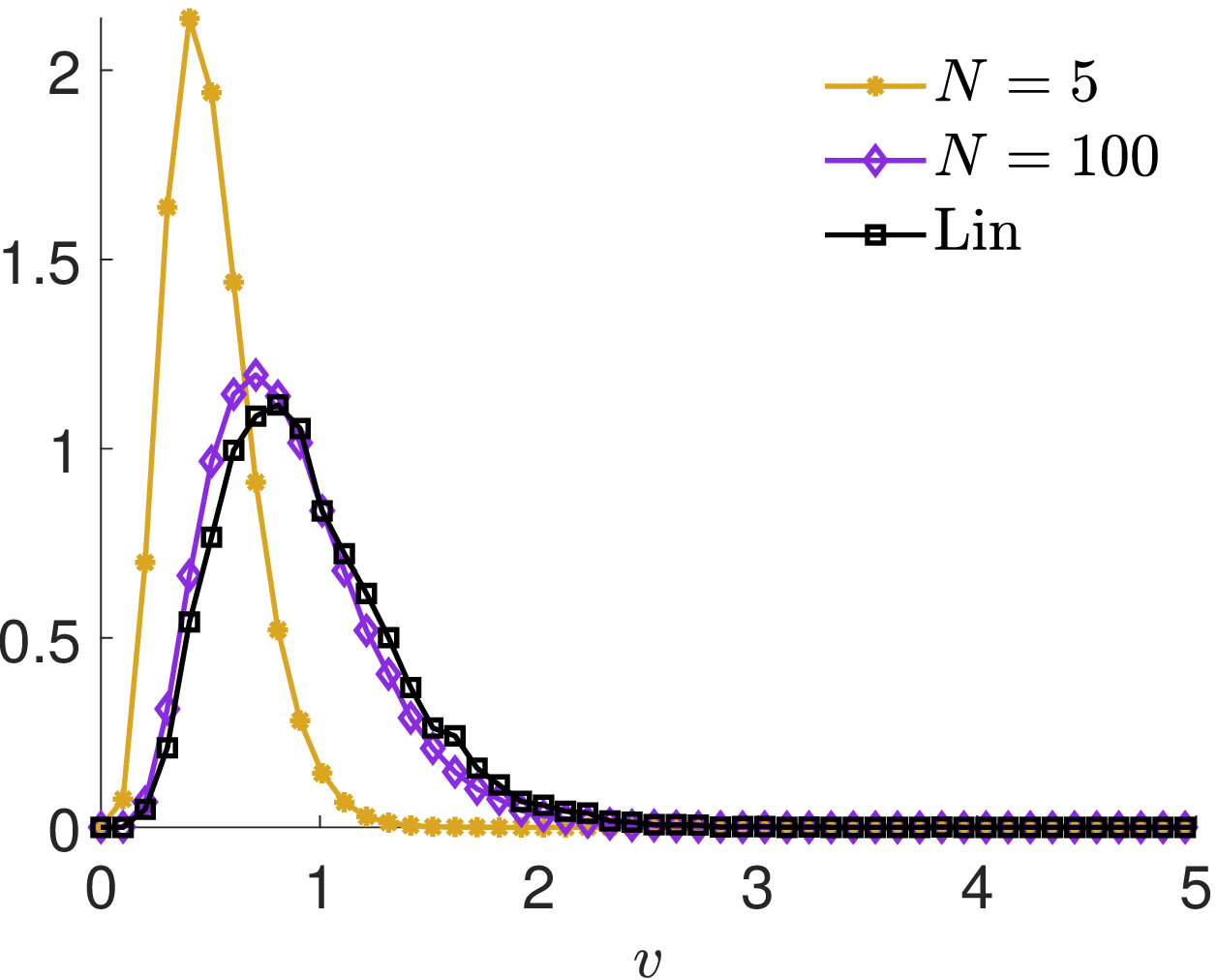}}
\caption{Evolution at times $t=1,\,5,\,25$ of the multiple-interaction model~\eqref{eq:N-int.gambling},~\eqref{eq:N-Boltz.gambling} with $N=5$, $N=100$ and of its linearised version~\eqref{eq:lin-int.gambling},~\eqref{eq:lin-Boltz.gambling}.}
\label{fig:gamb_1}
\end{figure}

In Figure~\ref{fig:gamb_1} we compare the evolution of the multiple-interaction model in the cases $N=5$, $N=100$ and that of the linearised model. We observe that if $N$ is sufficiently large then the solution of the former model is well approximated by that of the latter model. Conversely, for $N$ small strong differences appear, meaning that in such a case the multiple-interaction dynamics are quite poorly approximated by the linearised ones.

\begin{figure}[!t]
\centering
\includegraphics[scale=0.38]{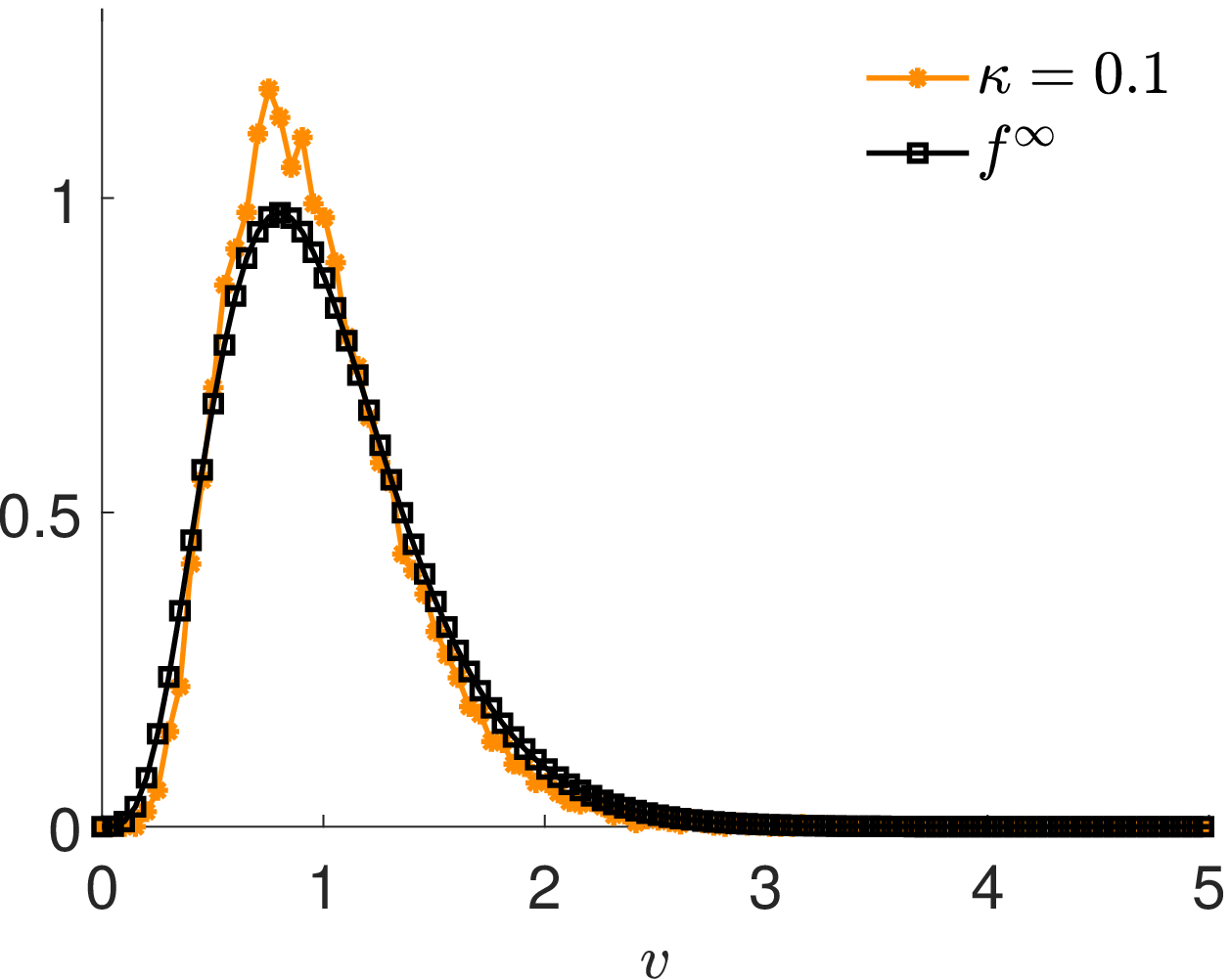}
\includegraphics[scale=0.38]{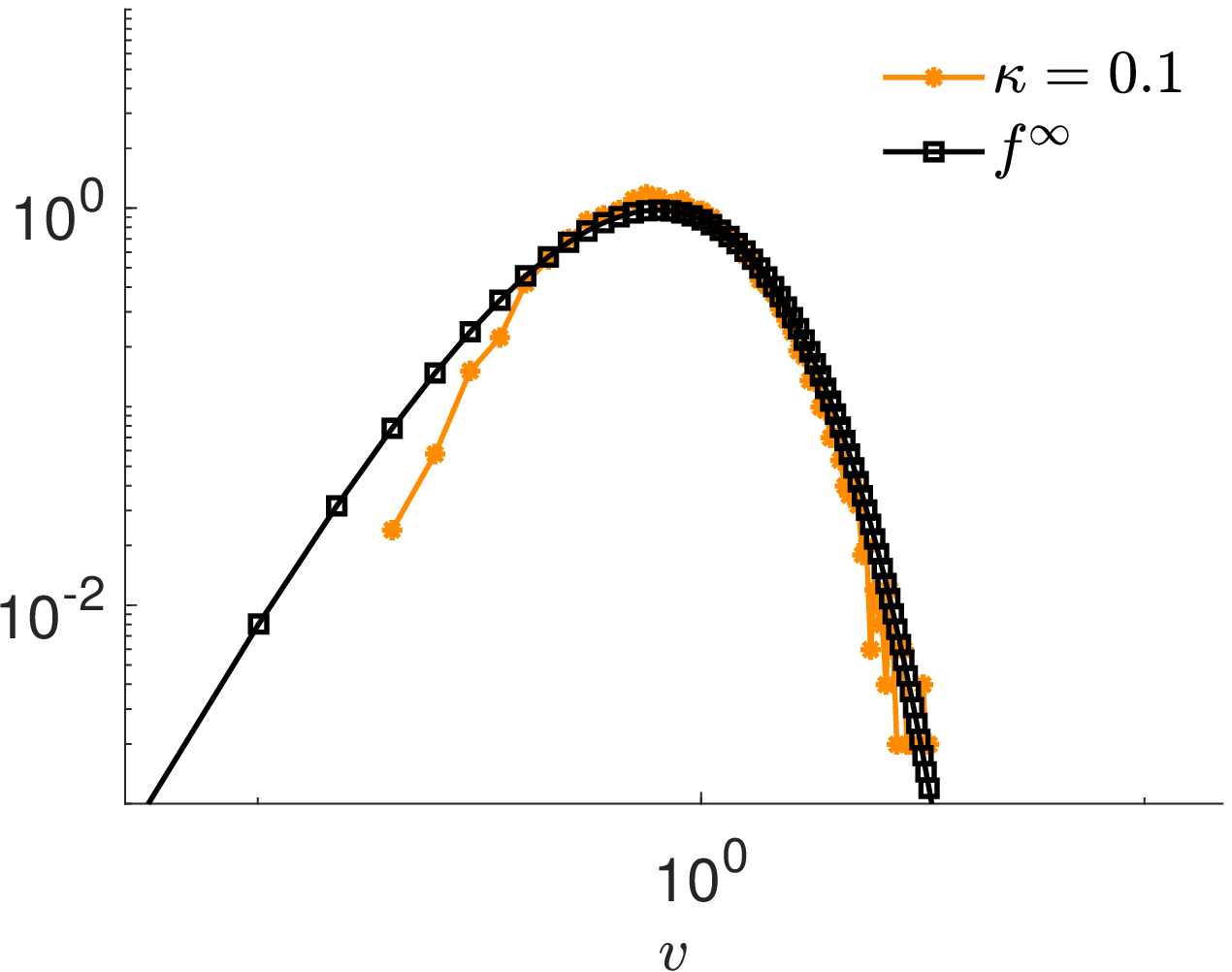} \\
\includegraphics[scale=0.38]{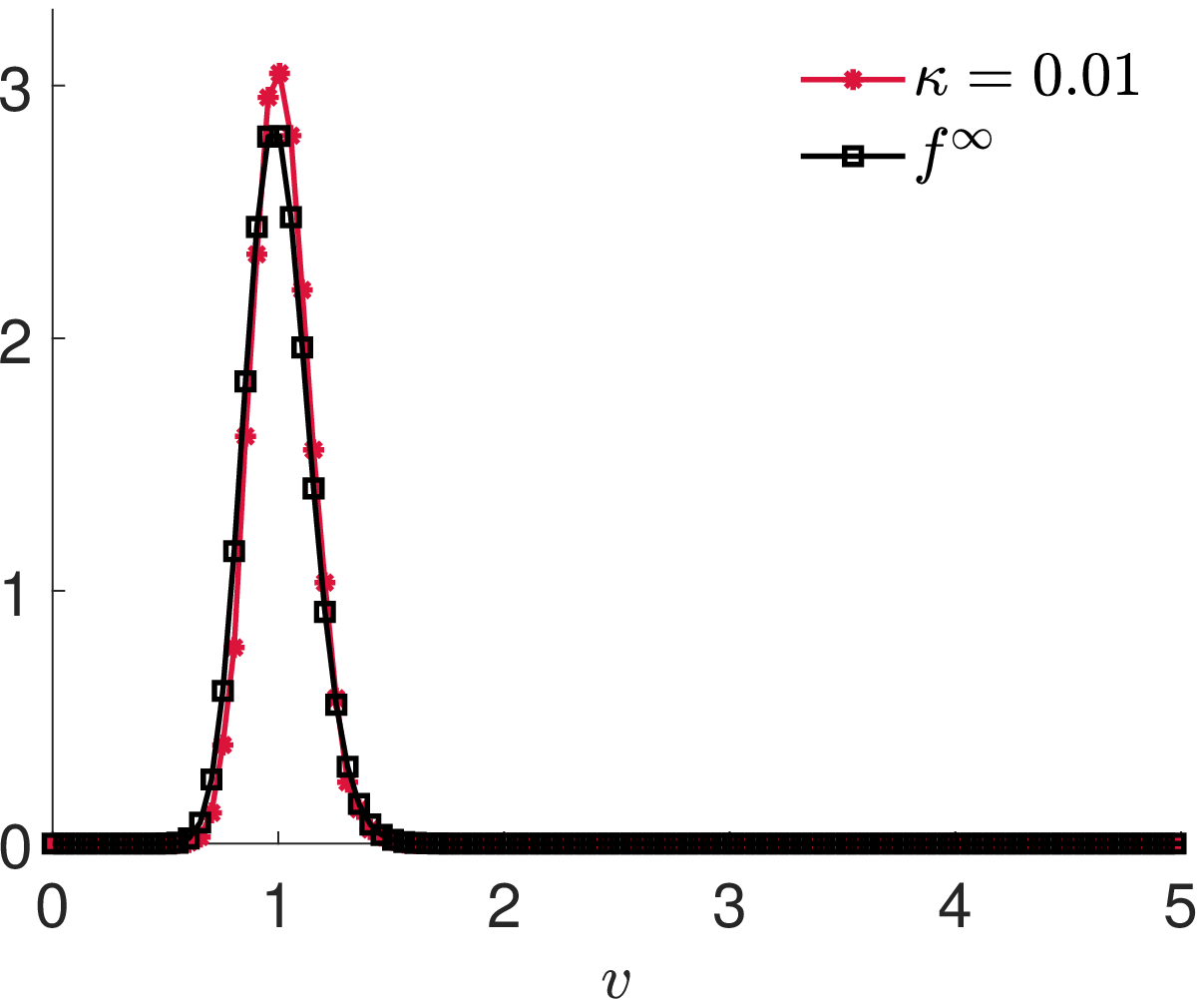}
\includegraphics[scale=0.38]{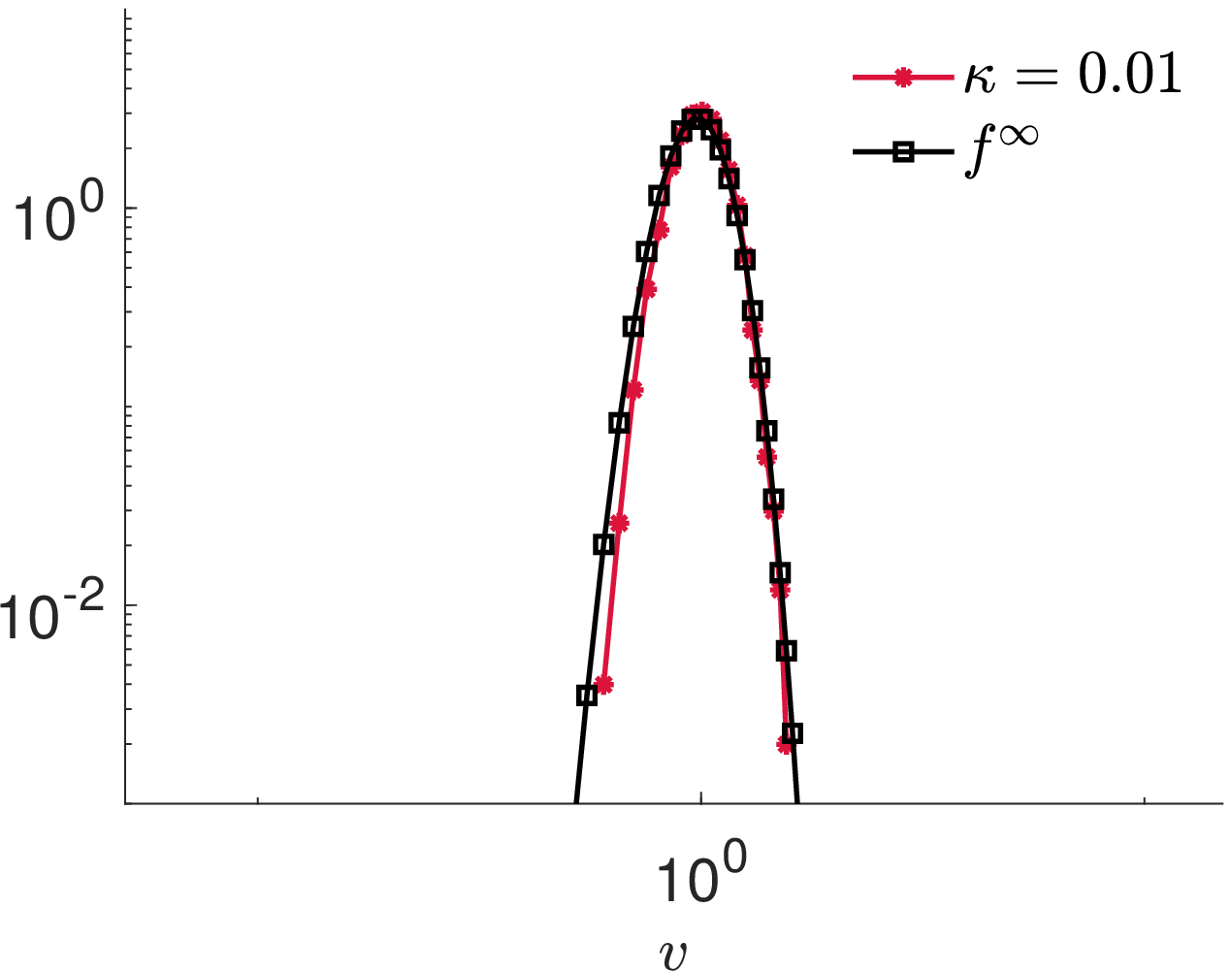}
\caption{Comparison between the large time solution ($t=25$) of the linearised model~\eqref{eq:lin-int.gambling},~\eqref{eq:lin-Boltz.gambling} and the equilibrium distribution $f^\infty$~\eqref{eq:gamma} computed from the Fokker-Planck equation~\eqref{eq:FP.gambling} in the quasi-invariant regime. \textbf{Top row}: $\kappa=0.1$, \textbf{bottom row}: $\kappa=0.01$. The right column displays the log-log plots of the graphs in the left column.}
\label{fig:gamb_2}
\end{figure}

In Figure~\ref{fig:gamb_2} we show instead the consistency between the linear model~\eqref{eq:lin-int.gambling},~\eqref{eq:lin-Boltz.gambling} and the Fokker-Planck asymptotic model~\eqref{eq:FP.gambling} obtained in the quasi-invariant limit and for small values of $\kappa$. In particular, we consider $\kappa=0.1$ and $\kappa=0.01$.

\begin{figure}[!t]
\centering
\includegraphics[scale=0.38]{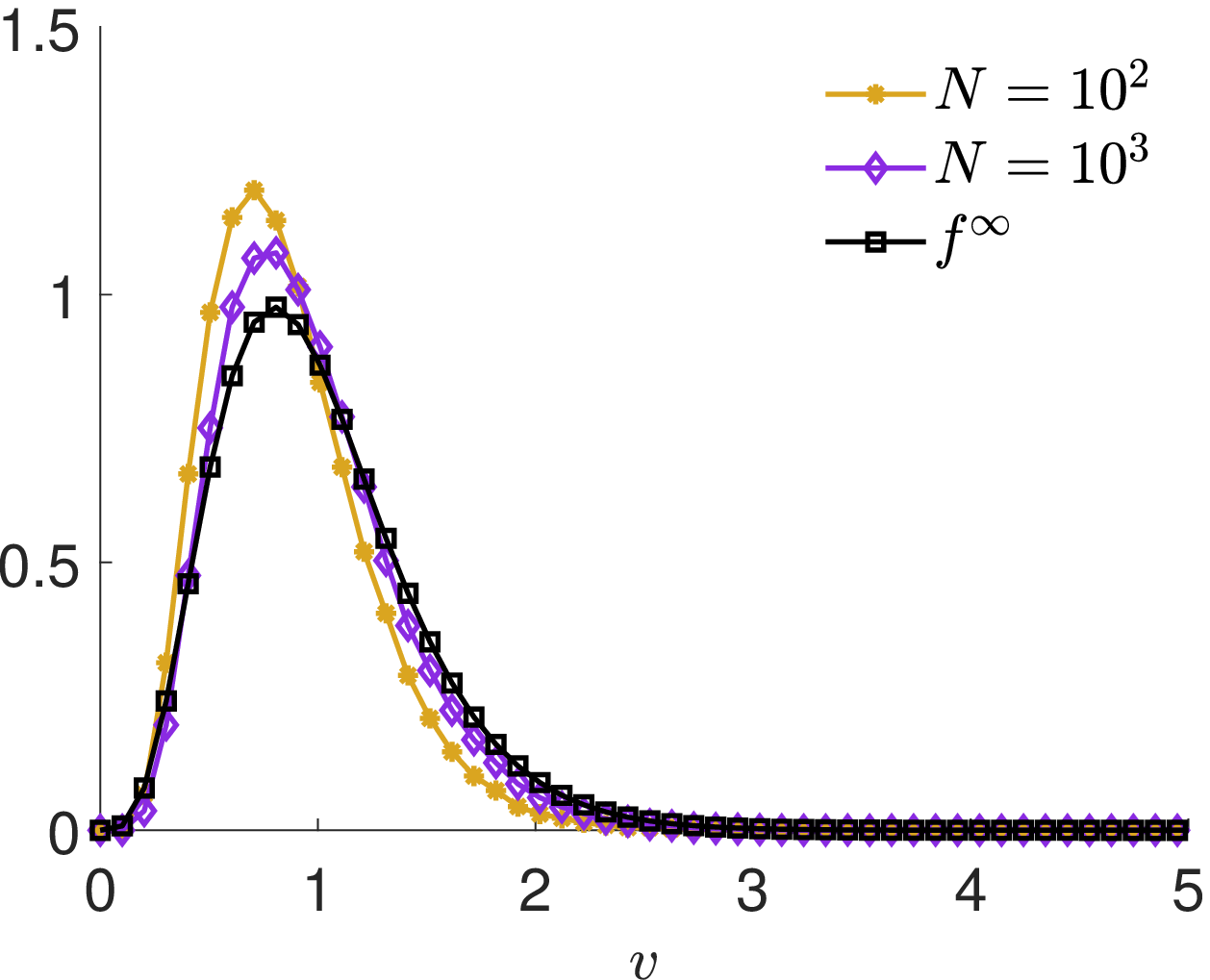} \qquad
\includegraphics[scale=0.38]{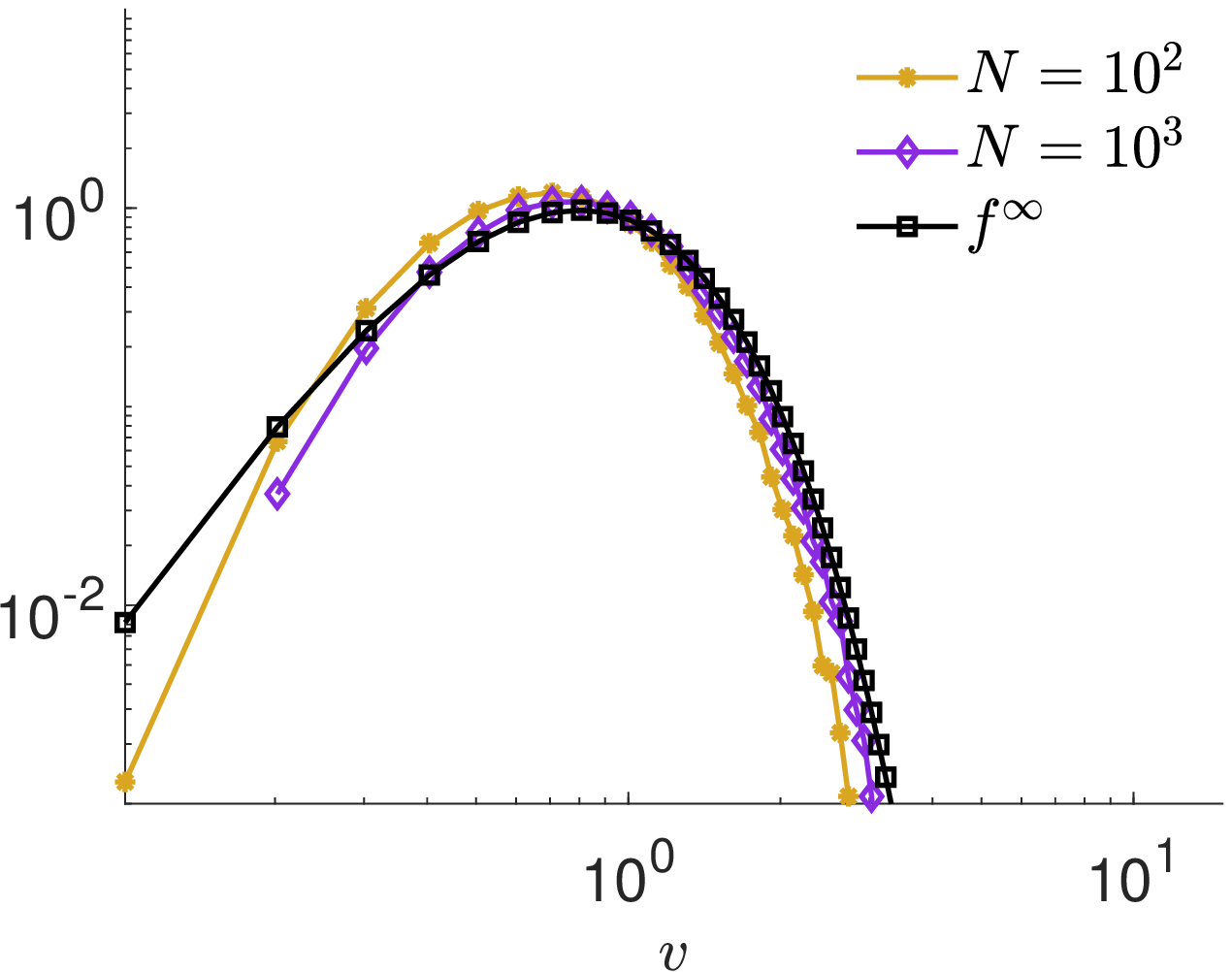}
\caption{Comparison between the equilibrium distribution $f^\infty$~\eqref{eq:gamma} and the large time solution of the multiple-interaction model with $N=10^2$ and $N=10^3$ for $\kappa=0.1$. The right panel is the log-log plot of the graph in the left panel, which gives a closer insight into the tails of the compared distributions.}
\label{fig:gamb_3}
\end{figure}

Finally, in Figure~\ref{fig:gamb_3} we compare the large time distribution of the multiple-interaction model for increasing numbers of gamblers and the steady distribution~\eqref{eq:gamma} computed from the Fokker-Planck asymptotic approximation of the linearised model. We fix $\kappa=0.1$ and we consider $N=10^2$, $N=10^3$. We clearly observe that for $N$ large enough the linearised model approximates well the multiple-interaction model also in the quasi-invariant regime.

\subsection{Taxation and wealth redistribution}
The second series of tests is devoted to the taxation and wealth redistribution model presented in Section~\ref{sect:taxation}. To describe statistically the background we choose a uniform distribution with probability density function
$$ g(w)=\chi_{[\frac{1}{10},\,\frac{11}{10}]}(w), $$
whereby the mean wealth of the background is $m=0.6$. Furthermore, we fix the taxation rate $\alpha=0.1$ and we choose $\lambda=\frac{1}{2(1+\alpha)}=0.3125$, which, according to the analysis of Section~\ref{sect:int_analysis.taxation}, guarantees the admissibility of the interactions~\eqref{eq:N-int.taxation}. Finally, we model the random variables $\eta_i$ as uniformly distributed with zero mean and variance equal to $\frac{\lambda}{10}$, cf.~\eqref{eq:eta_i}.

\begin{figure}[!t]
\centering
\includegraphics[scale=0.38]{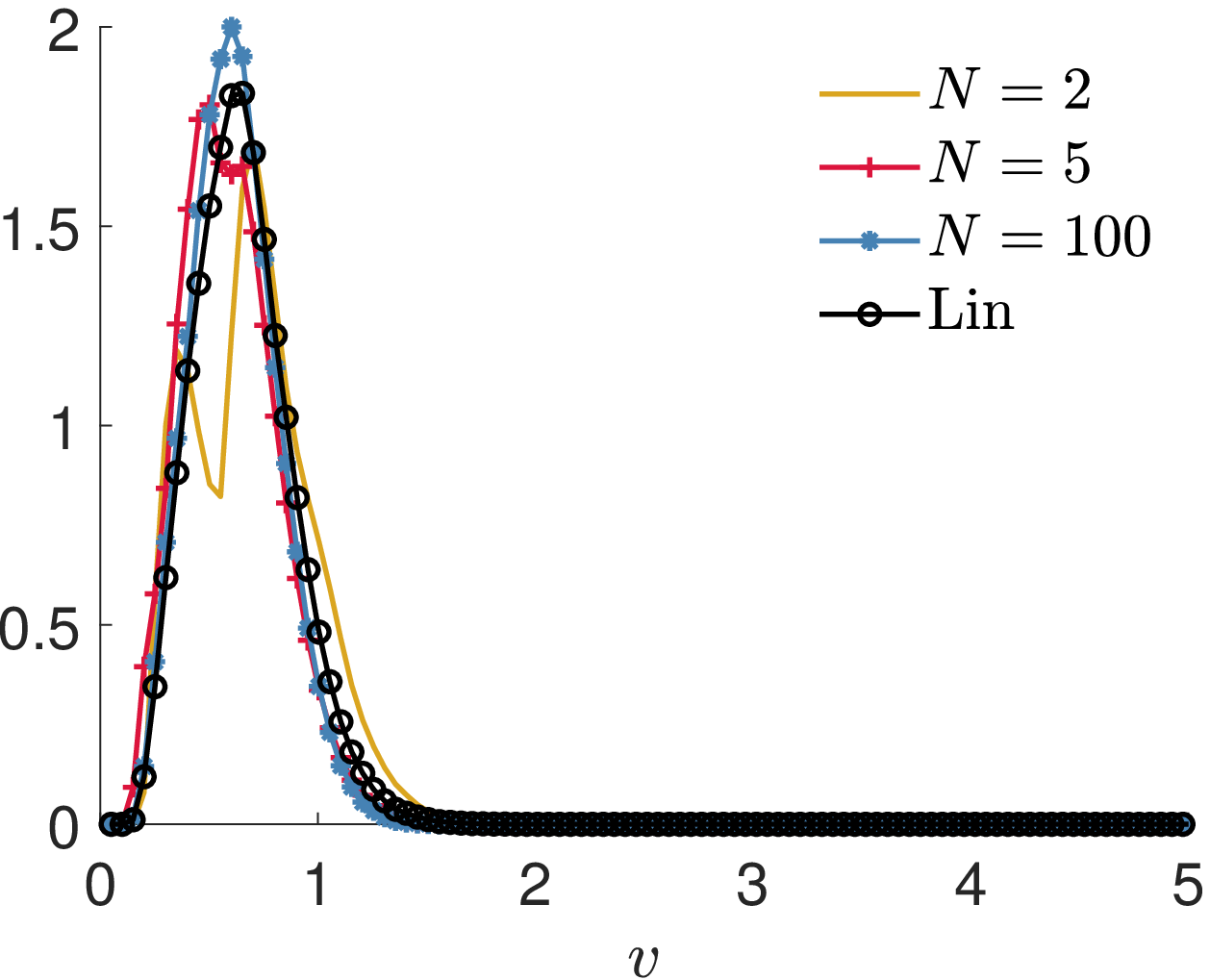}
\includegraphics[scale=0.38]{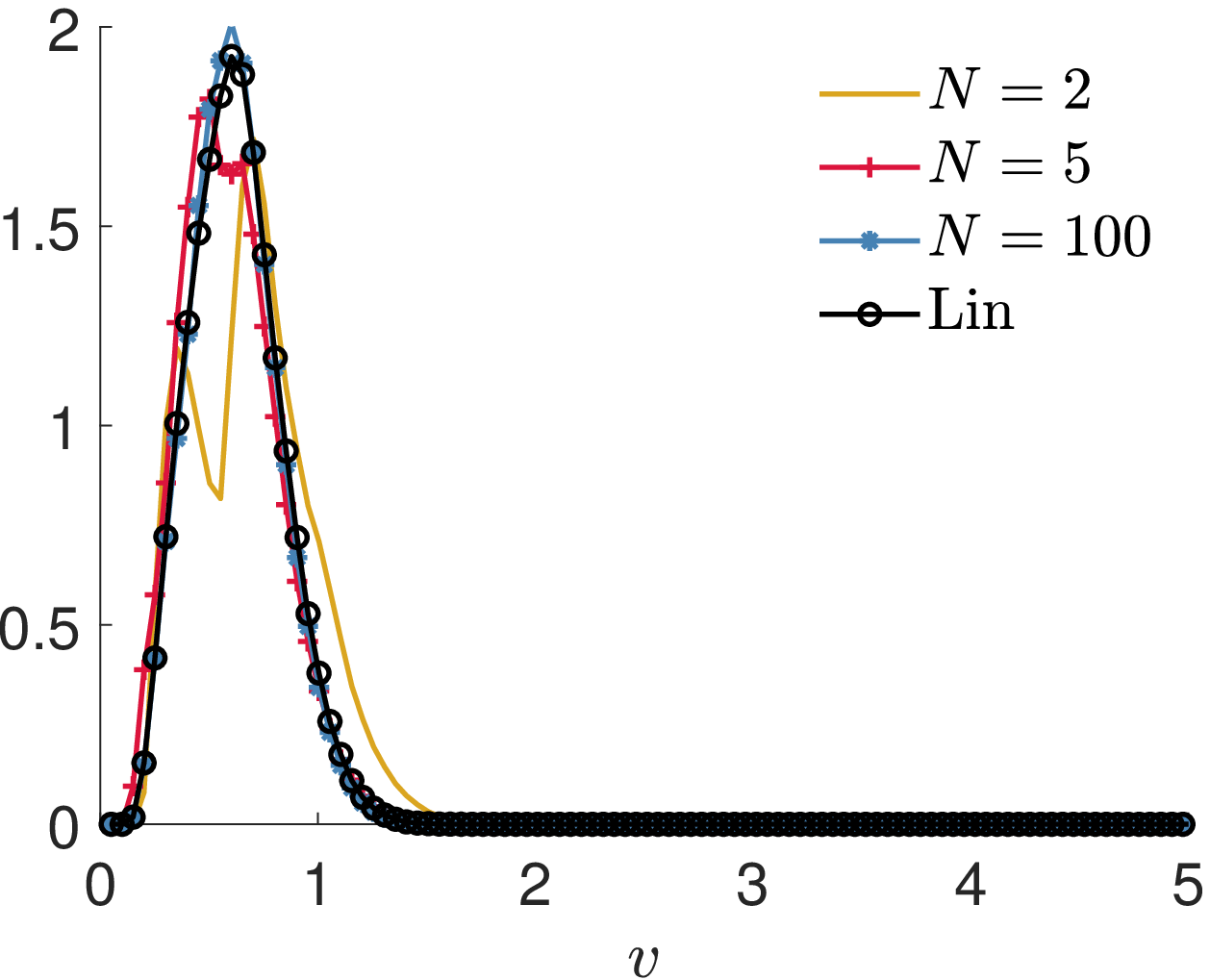}
\includegraphics[scale=0.38]{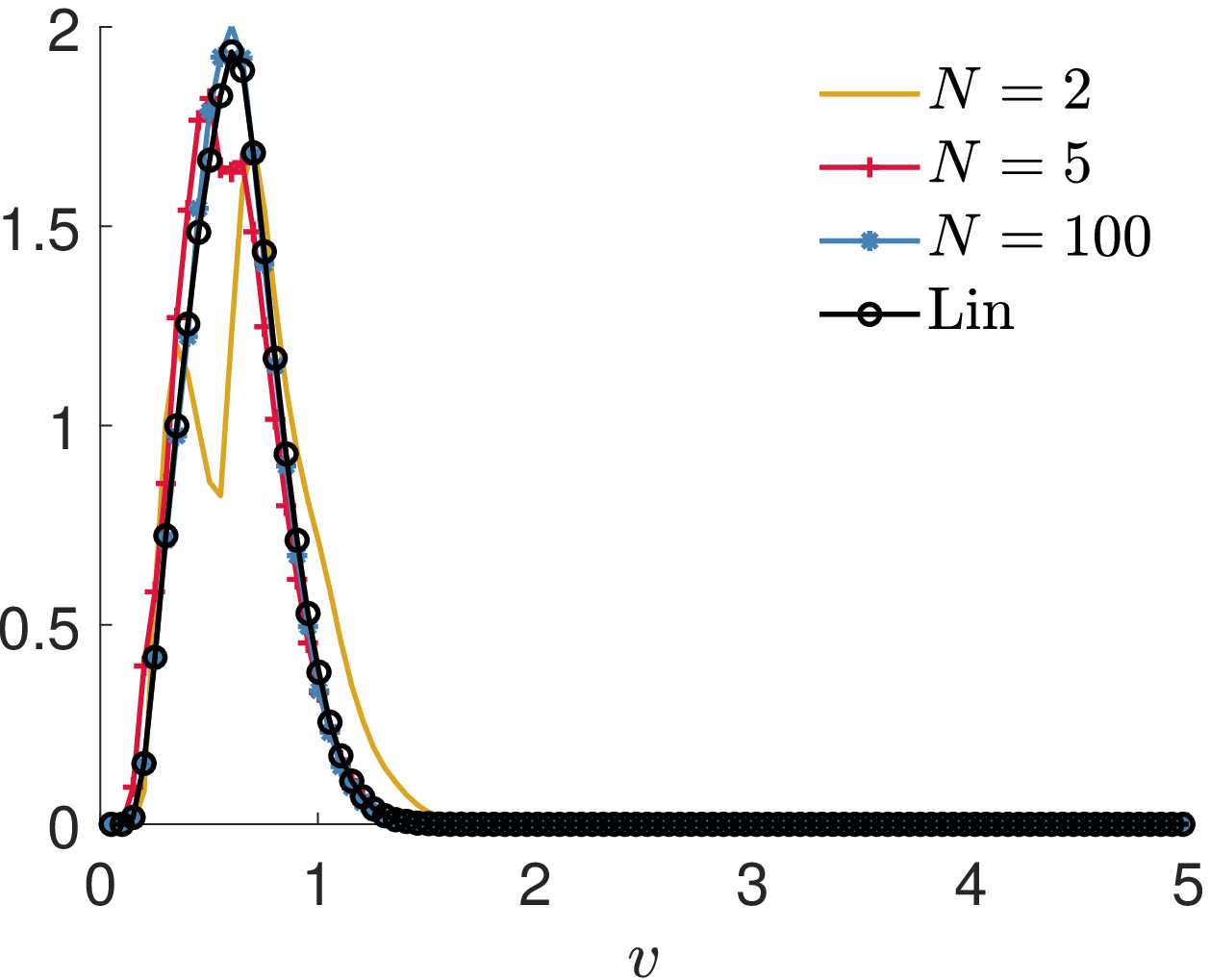} \\
\subfigure[$t=5$]{\includegraphics[scale=0.38]{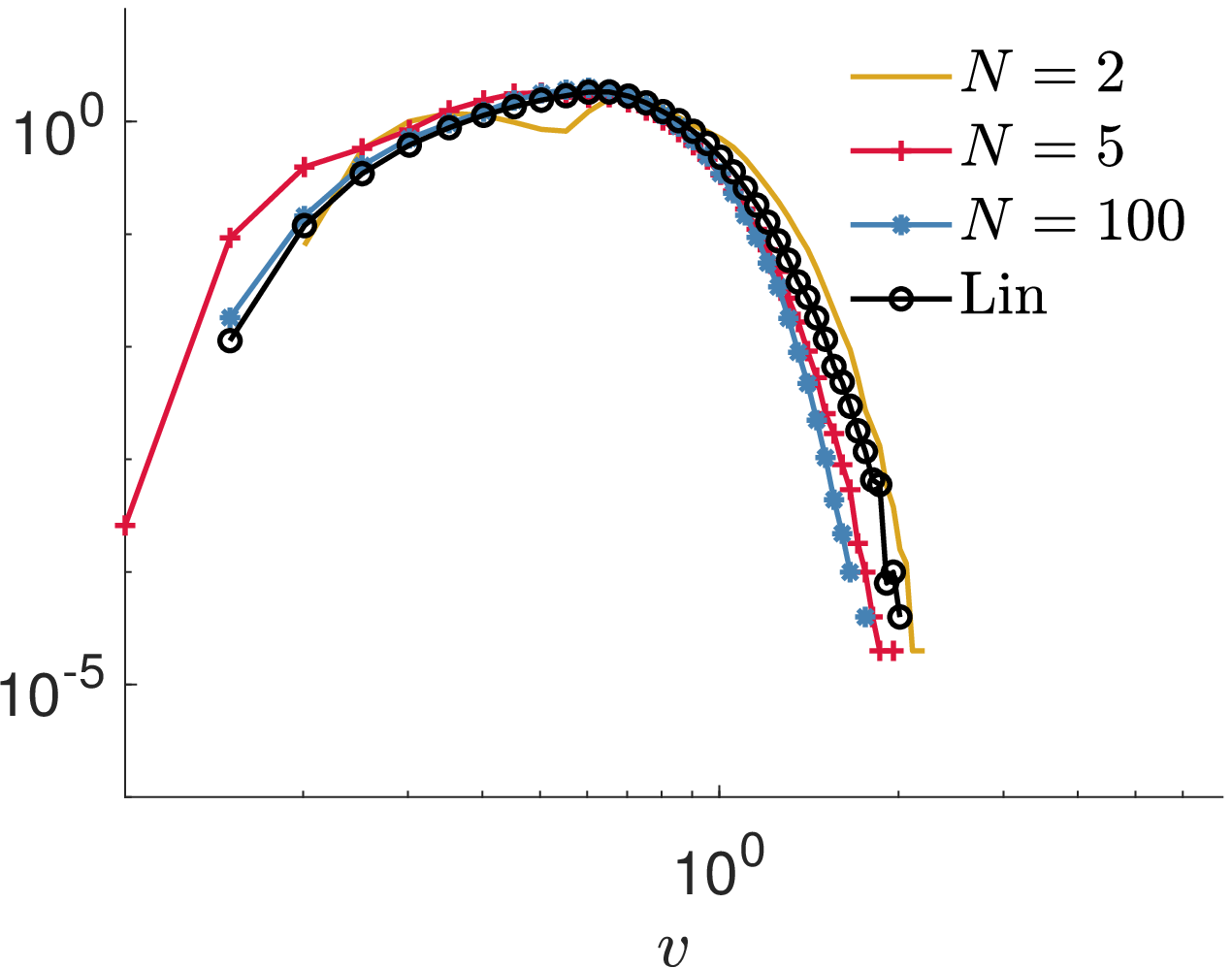}}
\subfigure[$t=10$]{\includegraphics[scale=0.38]{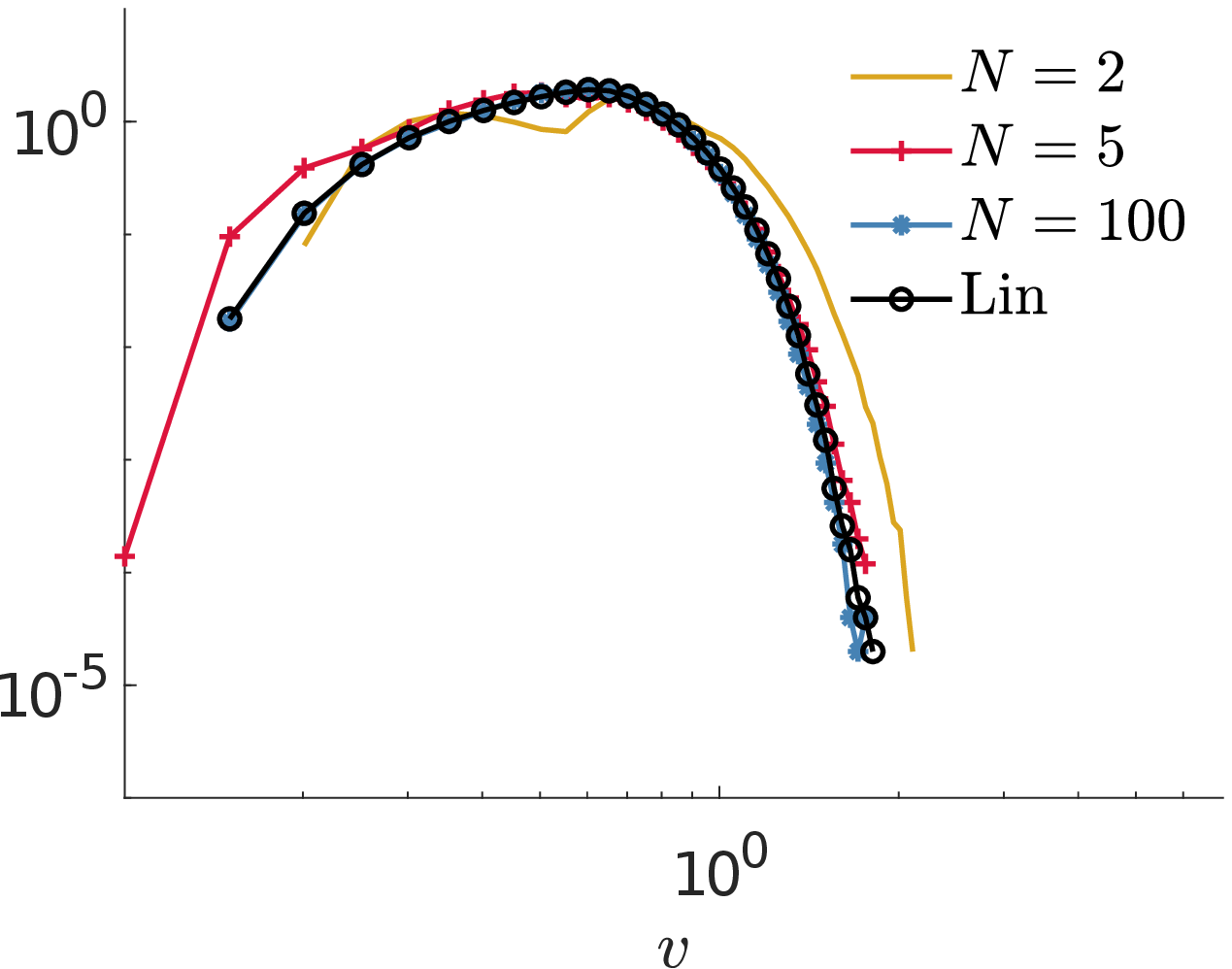}}
\subfigure[$t=25$]{\includegraphics[scale=0.38]{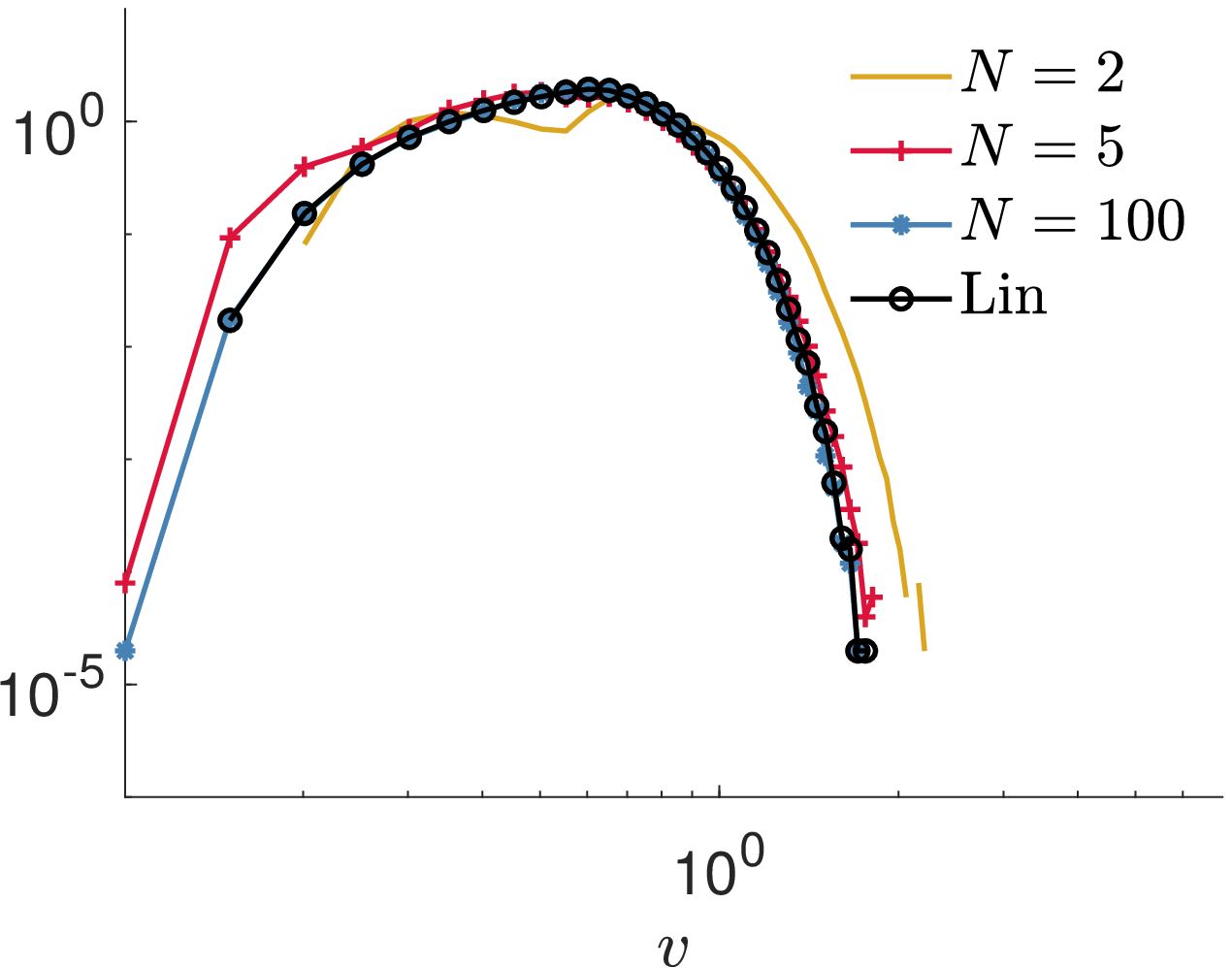}}
\caption{Evolution at times $t=5,\,10,\,25$ of the multiple-interaction model~\eqref{eq:N-int.taxation},~\eqref{eq:N-Boltz.taxation} with $N=2$, $N=5$, $N=100$ and of its linearised version~\eqref{eq:lin-int.taxation},~\eqref{eq:lin-Boltz.taxation}. The bottom row displays the log-log plots of the graphs in the top row to better appreciate the approximation of the tail of the distribution.}
\label{fig:tax1}
\end{figure}

Starting from the initial condition~\eqref{eq:f0}, in Figure~\ref{fig:tax1} we compare the evolution of the multiple-interaction model~\eqref{eq:N-int.taxation},~\eqref{eq:N-Boltz.taxation} with $N=2$, $N=5$, $N=100$ and that of its linearised version~\eqref{eq:lin-int.taxation},~\eqref{eq:lin-Boltz.taxation}. Also in this case, we observe that for increasing $N$ the linearisation produces a consistent approximation of the more complex multiple-interaction dynamics. Interestingly, a value of $N$ as small as $N=100$ appears to be sufficient for the linearised model to be a good approximation of the multiple-interaction model.

\begin{figure}[!t]
\centering
\includegraphics[scale=0.38]{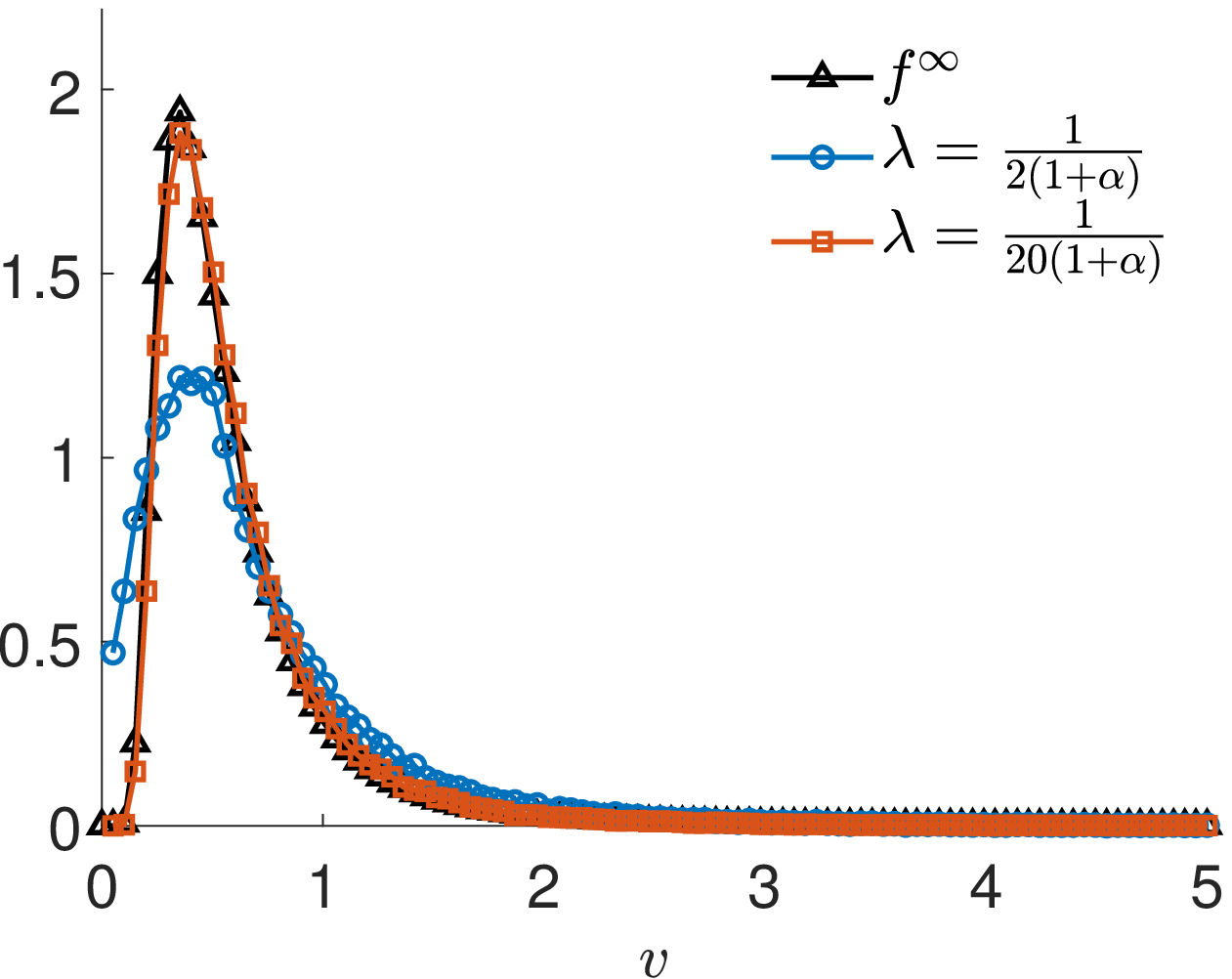} \qquad
\includegraphics[scale=0.38]{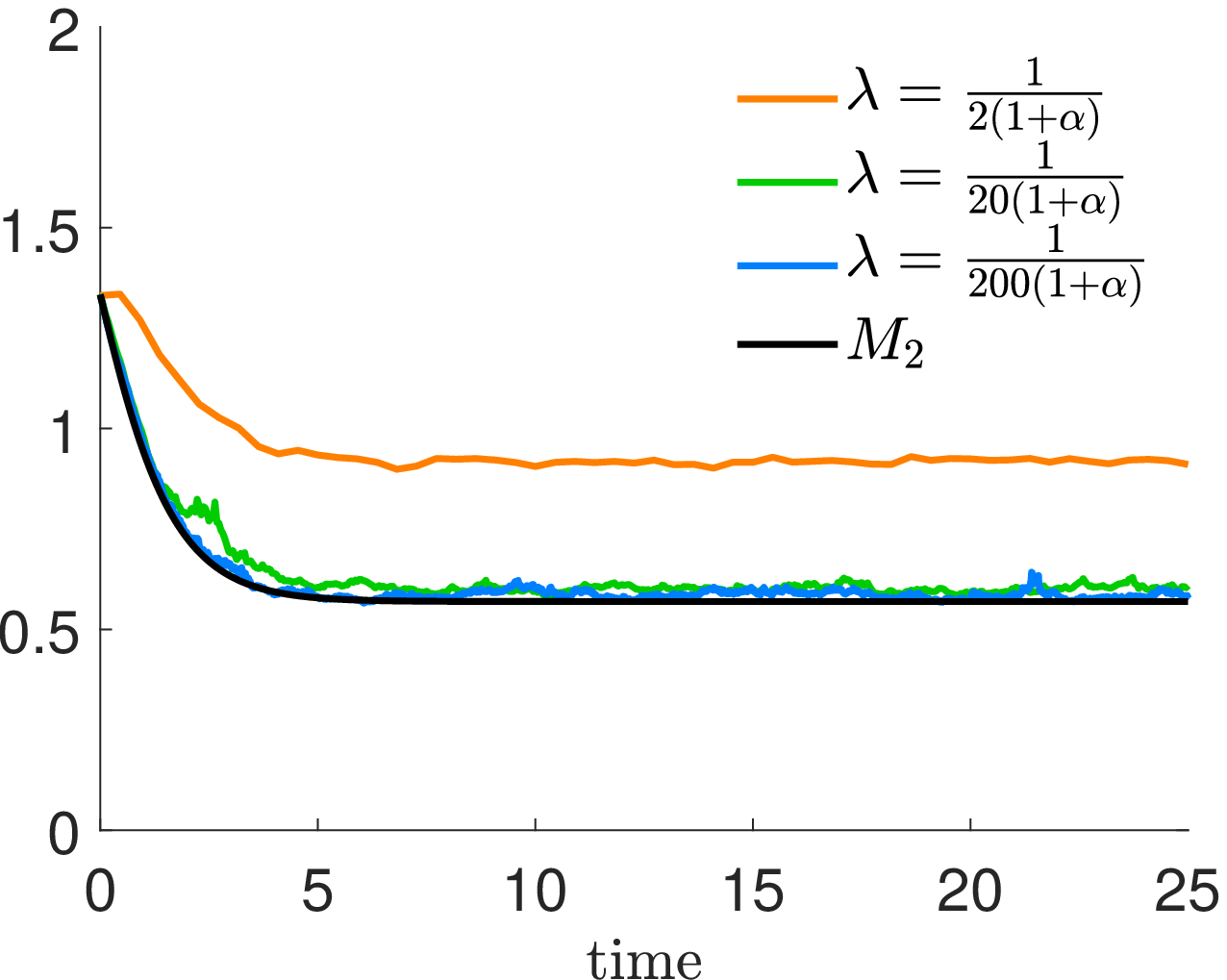}
\caption{\textbf{Left}: comparison of the equilibrium distribution~\ref{eq:inv_gamma} and the large time distribution of the linearised model~\eqref{eq:lin-int.taxation},~\eqref{eq:lin-Boltz_eps.taxation} in the quasi-invariant limit. \textbf{Right}: comparison of the evolution of the energy of the multiple-interaction model~\eqref{eq:N-Boltz.taxation} with $N=10$ for decreasing $\lambda$ and the solution to~\eqref{eq:tax_energy} obtained in the quasi-invariant limit.}
\label{fig:tax2}
\end{figure}

Moreover, in the left panel of Figure~\ref{fig:tax2} we show the convergence of the large time solution of the linearised model~\eqref{eq:lin-int.taxation},~\eqref{eq:lin-Boltz.taxation} to the equilibrium distribution $f^\infty$~\eqref{eq:inv_gamma} produced by the Fokker-Planck approximation~\eqref{eq:FP.taxation} in the quasi-invariant limit. In order to mimic such a small parameter regime we set $\tau=\lambda$ and we consider decreasing values of $\lambda$ of the form $\lambda=\frac{1}{2(1+\alpha)k}$ with $k=1,\,10$, so as to guarantee simultaneously the admissibility of the interaction~\eqref{eq:lin-int.taxation}.

Finally, to check in particular the convergence of the energy of the multiple-interaction model to the solution of~\eqref{eq:tax_energy} in the quasi-invariant regime, in the right panel of Figure~\ref{fig:tax2} we plot the time evolution of $M_2$ computed out of the numerical solution of the multiple-interaction model~\eqref{eq:N-int.taxation},~\eqref{eq:N-Boltz.taxation} with $N=10$ and several decreasing values of $\lambda$ of the form $\lambda=\frac{1}{2(1+\alpha)k}$ with $k=1,\,10,\,100$. Like before, this choice mimics the quasi-invariant limit. Furthermore, we also plot the numerical solution of system~\eqref{eq:quasi-inv.M1}-\eqref{eq:tax_energy} for duly comparison. In all cases, the initial data are fixed to $M_1(0)=1$, $M_2(0)=\frac{4}{3}$.

\section{Conclusions}
\label{sect:conclusions}
In recent years, the kinetic theory has proved to be a flexible and powerful tool to describe social and economic phenomena~\cite{pareschi2013BOOK}. In various situations, the precise description requires to take into account that any individual interacts simultaneously with a large number of other individuals. This implies that multiple simultaneous interactions have to be taken into account in collision-like kinetic models. In this paper, we have presented the main contributions to this new and challenging research line, which are concerned with socio-economic aspects. In particular, we have discussed two main examples: a model of the economical aspects of online jackpot games and a simple model of taxation and redistribution policy. The common aspect of these models is that the multiple interaction mechanism, which in principle gives rise to a highly non-linear version of the Boltzmann-type kinetic equation, can be greatly simplified, through a suitable linearisation, in the limit of a large number of individuals participating in each interaction. Numerical evidences then help to justify, at least formally, the simplified models obtained in the limit, suggesting that such an asymptotic approximation is often quite reliable also for a number of simultaneously interacting agents as moderately large as $O(10^2)$.

The kinetic modelling presented in this paper can be extended to cover other social situations, in which the emerging macroscopic state is still determined by interactions among many individuals. Among others, relevant cases that fit into this class are linked to human activities in social networks~\cite{toscani2018PRE}, which are assuming more and more relevance in determining the behaviour of the individuals. This new form of collective activity is very recent, since online social networking sites reached a very high popularity only in the last decade, inducing more and more individuals to connect with others who share similar interests~\cite{kuss2011IJERPH}. In particular, similarly to the case of the online gambling~\cite{toscani2019PRE}, it has been noticed that there is often an abuse of insights of social networking sites~\cite{kuss2011IJERPH}. This makes it topical to introduce and discuss some of these aspects, in which the multiple-interaction dynamics are the relevant ones, with the aim of extracting information about the aggregate macroscopic behaviour of the individuals. Results will be presented in a companion paper.

\section*{Acknowledgements}
This research was partially supported by the Italian Ministry of Education, University and Research (MIUR) through the ``Dipartimenti di Eccellenza'' Programme (2018-2022) -- Department of Mathematics ``F. Casorati'', University of Pavia and Department of Mathematical Sciences ``G. L. Lagrange'', Politecnico di Torino (CUP: E11G18000350001) and through the PRIN 2017 project (No. 2017KKJP4X) ``Innovative numerical methods for evolutionary partial differential equations and applications''.

This work is also part of the activities of the Starting Grant ``Attracting Excellent Professors'' funded by ``Compagnia di San Paolo'' (Torino) and promoted by Politecnico di Torino.

All the authors are members of GNFM (Gruppo Nazionale per la Fisica Matematica) of INdAM (Istituto Nazionale di Alta Matematica), Italy.

\bibliographystyle{plain}
\bibliography{TgTaZm-multiple}

\begin{thebibliography}{10}

\bibitem{bisi2017BUMI}
M.~Bisi.
\newblock Some kinetic models for a market economy.
\newblock {\em Boll. Unione Mat. Ital.}, 10(1):143--158, 2017.

\bibitem{bisi2009CMS}
M.~Bisi, G.~Spiga, and G.~Toscani.
\newblock Kinetic models of conservative economies with wealth redistribution.
\newblock {\em Commun. Math. Sci.}, 7(4):901--916, 2009.

\bibitem{bobylev2009CMP}
A.~V. Bobylev, C.~Cercignani, and I.~Gamba.
\newblock On the self-similar asymptotics for generalized nonlinear kinetic
  {M}axwell models.
\newblock {\em Comm. Math. Phys.}, 291(3):599--644, 2009.

\bibitem{bobylev2011KRM}
A.~V. Bobylev and \AA. Windfall.
\newblock Kinetic modeling of economic games with large number of participants.
\newblock {\em Kinet. Relat. Models}, 4(1):169--185, 2011.

\bibitem{brugna2018PHYSA}
C.~Brugna and G.~Toscani.
\newblock Kinetic models for goods exchange in a multi-agent market.
\newblock {\em Phys. A}, 499:362--375, 2018.

\bibitem{cordier2005JSP}
S.~Cordier, L.~Pareschi, and G.~Toscani.
\newblock On a kinetic model for a simple market economy.
\newblock {\em J. Stat. Phys.}, 120(1):253--277, 2005.

\bibitem{dimarco2014AN}
G.~Dimarco and L.~Pareschi.
\newblock Numerical methods for kinetic equations.
\newblock {\em Acta Numer.}, 23:369--520, 2014.

\bibitem{dimarco2019JSP}
G.~Dimarco and G.~Toscani.
\newblock Kinetic modeling of alcohol consumption.
\newblock {\em J. Stat. Phys.}, In press.
\newblock Preprint: arXiv:1902.08198.

\bibitem{duering2018EPJB}
B.~D\"{u}ring, L.~Pareschi, and G.~Toscani.
\newblock Kinetic models for optimal control of wealth inequalities.
\newblock {\em Eur. Phys. J. B}, 91:265/1--12, 2018.

\bibitem{ernst2002JSP}
M.~H. Ernst and R.~Brito.
\newblock Scaling solutions of inelastic {B}oltzmann equations with
  over-populated high energy tails.
\newblock {\em J. Stat. Phys.}, 109(3--4):407--432, 2002.

\bibitem{garibaldi2007EPJB}
U.~Garibaldi, E.~Scalas, and P.~Viarengo.
\newblock Statistical equilibrium in simple exchange games {II}. {T}he
  redistribution game.
\newblock {\em Eur. Phys. J. B}, 60(2):241--246, 2007.

\bibitem{guala2009INDECS}
S.~Guala.
\newblock Taxes in a wealth distribution model by inelastically scattering of
  particles.
\newblock {\em Interdisciplinary Description of Complex Systems}, 7(1):1--7,
  2009.

\bibitem{gualandi2018ECONOMICS}
S.~Gualandi and G.~Toscani.
\newblock {P}areto tails in socio-economic phenomena: a kinetic description.
\newblock {\em Economics}, 12(2018-31):1--17, 2018.

\bibitem{gualandi2019M3AS}
S.~Gualandi and G.~Toscani.
\newblock Human behavior and lognormal distribution. a kinetic description.
\newblock {\em Math. Models Methods Appl. Sci.}, 29(4):717--753, 2019.

\bibitem{kuss2011IJERPH}
D.~J. Kuss and M.~D. Griffiths.
\newblock Online social networking and addiction -- {A} review of the
  physchological literature.
\newblock {\em Int. J. Environ. Res. Public Health}, 8(9):3528--3552, 2011.

\bibitem{pareschi2013BOOK}
L.~Pareschi and G.~Toscani.
\newblock {\em Interacting {M}ultiagent {S}ystems: {K}inetic equations and
  {M}onte {C}arlo methods}.
\newblock Oxford University Press, 2013.

\bibitem{pareschi2018JSC}
L.~Pareschi and M.~Zanella.
\newblock Structure preserving schemes for nonlinear {F}okker-{P}lanck
  equations and applications.
\newblock {\em J. Sci. Comput.}, 74(3):1575--1600, 2018.

\bibitem{slanina2004PRE}
F.~Slanina.
\newblock Inelastically scattering particles and wealth distribution in an open
  economy.
\newblock {\em Phys. Rev. E}, 69(4):046102/1--7, 2004.

\bibitem{toscani2006CMS}
G.~Toscani.
\newblock Kinetic models of opinion formation.
\newblock {\em Commun. Math. Sci.}, 4(3):481--496, 2006.

\bibitem{toscani2009EPL}
G.~Toscani.
\newblock Wealth redistribution in conservative linear kinetic models.
\newblock {\em Europhys. Lett. EPL}, 88(1), 2009.

\bibitem{toscani2013JSP}
G.~Toscani, C.~Brugna, and S.~Demichelis.
\newblock Kinetic models for the trading of goods.
\newblock {\em J. Stat. Phys.}, 151(3-4):549--566, 2013.

\bibitem{toscani2018PRE}
G.~Toscani, A.~Tosin, and M.~Zanella.
\newblock Opinion modeling on social media and marketing aspects.
\newblock {\em Phys. Rev. E}, 98(2):022315/1--15, 2018.

\bibitem{toscani2019PRE}
G.~Toscani, A.~Tosin, and M.~Zanella.
\newblock Multiple-interaction kinetic modeling of a virtual-item gambling
  economy.
\newblock {\em Phys. Rev. E}, 100(1):012308/1--16, 2019.

\bibitem{wang2018PRE}
X.~Wang and M.~Pleimling.
\newblock Behavior analysis of virtual-item gambling.
\newblock {\em Phys. Rev. E}, 98:012126/1--12, 2018.

\bibitem{Wang2019}
X.~Wang and M.~Pleimling.
\newblock Online gambling of pure chance: {W}ager distribution, risk attitude,
  and anomalous diffusion.
\newblock {\em Sci. Rep.}, 9:14712, 2019.

\end{thebibliography}
\end{document}